\documentclass[a4paper,fleqn,usenatbib]{mnras}

\usepackage{mathptmx}

\usepackage[T1]{fontenc}
\usepackage{ae,aecompl}

\usepackage{booktabs}
\usepackage{xtab}
\usepackage{multirow}

\usepackage{graphicx}	
\usepackage{amsmath}	
\usepackage{amssymb}	

\newcommand{\ltsima}{$\; \buildrel < \over \sim \;$}
\newcommand{\simlt}{\lower.5ex\hbox{\ltsima}}
\newcommand{\gtsima}{$\; \buildrel > \over \sim \;$}
\newcommand{\simgt}{\lower.5ex\hbox{\gtsima}}

\newcommand{\cgs}{${\rm erg~cm}^{-2}~{\rm s}^{-1}$} 
 
\newcommand{\lum}{\rm erg~s$^{-1}$}

\newcommand{\pn}{\par\noindent}

\def\lesssim{\mathrel{\hbox{\rlap{\hbox{\lower4pt\hbox{$\sim$}}}\hbox{$<$}}}}
\def\gtrsim{\mathrel{\hbox{\rlap{\hbox{\lower4pt\hbox{$\sim$}}}\hbox{$>$}}}}

\def\arcdeg{\hbox{$^\circ$}}

\def\arcsec{\hbox{$^{\prime\prime}$}}
\def\micron{\hbox{$\mu$m}}

\def\ab1450{$AB_{1450(1+z)}$}

\def\xray{\hbox{X-ray}}

\def\09104{IRAS~09104$+$4109}
\def\I09104{I09104}

\def\edd_ratio{$\log\ L_{\rm bol}/L_{\rm Edd}$}

\def\l58{{$(\lambda L_{\lambda})_{\mbox{{\rm \scriptsize 5.8\micron}}}$}}
\def\lmir2{{$(\lambda L_{\lambda})_{\mbox{{\rm \scriptsize 12.3\micron}}}$}}
\def\s1{{S$_{\mbox{{\rm \scriptsize 3.6\micron}}}$}}
\def\irac2{{S$_{\mbox{{\rm \scriptsize 4.5\micron}}}$}}
\def\f3{{S$_{\mbox{{\rm \scriptsize 5.8\micron}}}$}}
\def\mic8{{S$_{\mbox{{\rm \scriptsize 8\micron}}}$}}
\def\f24{{F$_{\mbox{{\rm \scriptsize 24\micron}}}$}}

\def\chandra{{\it Chandra\/}}

\def\heao1{{\it HEAO-1\/}}

\def\iras{{\it IRAS\/}}

\def\spitzer{{\it Spitzer\/}}

\def\nustar{{\it NuSTAR\/}}

\def\xmm{{XMM-{\it Newton\/}}}
\def\suzaku{{\it Suzaku\/}}

\def\swift{{\it Swift\/}}
\def\integral{{\it Integral\/}}
\def\erosita{{\it eROSITA\/}}
\def\spica{{\it SPICA\/}}

\title[X-ray analysis of Compton-thick AGN]{Broad-band X-ray analysis of local mid-infrared selected Compton-thick AGN candidates}

\author[M.-M. La Caria et al.]{M.-M. La Caria,$^{1,2}$\thanks{E-mail: {marlismadele.lacaria@studio.unibo.it.}}
C. Vignali,$^{1,3}$
G. Lanzuisi,$^{3}$
C. Gruppioni$^{3}$
and F. Pozzi$^{1,3}$
\\
$^{1}$ Dipartimento di Fisica e Astronomia, Universit\`{a} di Bologna, Via Piero Gobetti 93/2, 40129 Bologna, Italy\\
$^{2}$ Max-Planck-Institut f{\"u}r Extraterrestriche Physik, Giessenbachstrasse, D-85748, Garching, Germany \\
$^{3}$ INAF--Osservatorio di Astrofisica e Scienza dello Spazio di Bologna, Via Piero Gobetti 93/3, 40129 Bologna, Italy
}

\date{Accepted 2019 May 14. Received 2019 May 6; in original form 2018 December 30.}

\defcitealias{Gruppioni2016}{G16}
\defcitealias{BrightmanNandra2011a}{BN11a}
\defcitealias{Marconi2004}{M04}
\defcitealias{Masini2016}{M16}
\defcitealias{Marchesi2018}{M18}

\begin{document}
\label{firstpage}
\pagerange{\pageref{firstpage}--\pageref{lastpage}}
\maketitle

\begin{abstract}
The estimate of the number and space density of obscured AGN over cosmic time still represents an open issue. While the obscured AGN population is a key ingredient of the X-ray background synthesis models and is needed to reproduce its shape, a complete census of obscured AGN is still missing. 
Here we test the selection of obscured sources among the local 12\micron\ sample of Seyfert galaxies. Our selection is based on a difference up to three orders of magnitude in the ratio between the AGN bolometric luminosity, derived from the spectral energy distribution (SED) decomposition, and the same quantity obtained by the published \xmm\ 2--10~keV luminosity ($L_{bol}^{AGN}(IR)/L_{bol}^{AGN}(X)$). 
The selected sources are UGC05101, NGC1194 and NGC3079 for which the available \xray\ wide bandpass, from \chandra\ and \xmm\ plus \nustar\ data, extending to energies up to $\sim$30--45~keV, allows us an accurate determination of the column density, and hence of the true intrinsic power. 
The newly derived $N_{H}$ values clearly indicate heavy obscuration ($\sim1.2$, $2.1$ and $2.4\times10^{24}$~cm$^{-2}$ for UGC05101, NGC1194 and NGC3079, respectively), and are consistent with the prominent silicate absorption feature observed in the \spitzer-IRS spectra of these sources (9.7\micron\ rest frame). We finally checked that the resulting \xray\ luminosities in the 2--10~keV band are in good agreement with those derived from the mid-IR band through empirical $L_{MIR}-L_{X}$ relations. 
\end{abstract}




\section{Introduction}

XRB synthesis models \citep[e.g.,][]{Gilli2007, Akylas2012} require a significant fraction of obscured active galactic nuclei (AGN); a fraction of these are expected to be heavily obscured, including Compton-thick AGN (with column density $N_{H}\geq 1.5\times10^{24}$~cm$^{-2}$), thus likely elusive or difficult to detect even in deep \xray\ surveys. 
The number density of obscured sources and their space density as a function of cosmic time still represent an open and challenging issue \citep[e.g.,][]{Vignali2010, Vignali2014}. 
Being hidden by extreme column densities of obscuring gas that can absorb even hard \xray\ photons, Compton-thick AGN are difficult to find, especially at high redshifts. Mildly Compton-thick AGN (column densities of $\sim10^{24}$--$10^{25}$~cm$^{-2}$) are the most promising candidates to explain the residual (i.e., not resolved yet into individual sources at the limits of current \xray\ surveys) spectrum of the cosmic XRB at its 30~keV peak \citep[e.g.,][]{Gilli2013, Shi2013} 
but only a limited number of these sources are currently known and verified as bona-fide Compton-thick AGN beyond the local Universe. 
The redshift distribution and space density of Compton-thick AGN are also important in accurately assessing the supermassive black hole mass function and constraining the Eddington ratio and radiative efficiency averaged over cosmic times; this result can be achieved through a comparison with the relic black hole mass distribution of local galaxies e.g., \citeauthor{Marconi2004} \textcolor{blue}{2004}, hereafter M04).
%

Obscured AGN look differently depending on the band in which they are observed. Therefore, to provide a census as complete as possible of these AGN, it is fundamental to follow a multi-wavelength approach and keep into consideration the observational biases behind each of the adopted source selection methods. 
The safest way to identify Compton-thick AGN and describe the basic \xray\ properties of their nuclear emission consists of detecting the primary continuum piercing through the absorber, but this typically requires measurements above 10~keV. Below 10~keV, the classification of a source as a Compton-thick AGN relies mostly on indirect indications, such as the observation of a flat photon index for the hard \xray\ continuum, the presence of a reflected and/or scattered continuum, a prominent iron K$\alpha$ fluorescence emission line (with typical equivalent width, EW, of $\geq1$~keV), and anomalously low values for the ratio between the observed \xray\ flux and that predicted on the basis of AGN intrinsic emission proxies at other wavelengths. 

Over the last fifteen years, in the deepest fields $>$80\% of the 2--10~keV XRB has been resolved into individual objects by \chandra\ and \xmm\ surveys \citep{Hickox2006, Brandt2015}. At energies above 10~keV, however, the observational framework is far from being complete, due to observational limitations. The sensitivity of coded-mask instruments such as \integral/IBIS and \swift/BAT ($\sim 10^{-11}$~\cgs) is such that this investigation is limited to the local Universe \citep[e.g.,][]{Tueller2008, Beckmann2009}; in the end, the fraction of the XRB resolved by these instruments at its peak intensity (20--30~keV) is $\sim1-2$\% \citep{Krivonos2007, Ajello2012, Vasudevan2013}. The field regarding hard \xray\ surveys has been revitalized by the hard \xray\ imaging instrument onboard \nustar, which has improved the sensitivity limits by a factor of more than 100 with respect to \integral\ and \swift, and has probed a significantly wider range in redshift, up to z$\sim$3 \citep{Lansbury2017a}; we note, however, that most of the \nustar-detected sources lie at low redshift. \nustar\ has now resolved $\sim$35\% of the XRB in the 8--24~keV band and directly identified AGN with high obscuring column densities \citep{Ricci2015, Harrison2016, Lansbury2017b, DelMoro2017, Zappacosta2018}. 

This work takes advantage of \nustar\ data and is based on a multi-wavelength research approach. The work carried out by \citeauthor{Gruppioni2016} (\textcolor{blue}{2016}, hereafter G16) on the local 12\micron\ sample of Seyfert galaxies \citep[12MGS;][]{Rush1993} with \spitzer-IRS spectra has played a key role in defining the selection criteria for the sample of galaxies investigated here. We used the infrared observations along with X-rays to pick up and properly investigate three local galaxies which are likely to host a heavily obscured AGN. The mid-infrared (mid-IR) regime offers a good potential for discoveries of heavily obscured AGN, since any primary AGN continuum (i.e., disc emission in the optical/UV) that is absorbed must ultimately come out at these wavelengths after being thermally reprocessed by the dusty nuclear material, i.e., the torus.
\citetalias{Gruppioni2016} compared the AGN bolometric luminosity ($L_{bol}^{AGN}$) derived from the SED decomposition (hereafter $L_{bol}^{AGN} (IR)$) to the same quantity obtained by means of different methods: the 2--10~keV luminosity (from IPAC-NED public database) and the AGN mid-IR line luminosity ([Ne V]14.3\micron, [Ne V]24.3\micron\ and 
[O IV]25.9\micron; \citealt{Tommasin2008, Tommasin2010}). As the \xray\ luminosities taken from literature and based on data below 10~keV were already corrected for absorption, $L_{bol}^{AGN} (X)$ (i.e., the bolometric luminosity derived from the \xray\ emission adopting a bolometric correction) should represent the intrinsic emission from the nucleus, and the $L_{bol}^{AGN}(IR)/L_{bol}^{AGN}(X)$ ratios should be close to unity. Instead, $L_{bol}^{AGN}(X)$ up to three orders of magnitude lower than $L_{bol}^{AGN}(IR)$ were found for some sources. One possibility is that the intrinsic \xray\ luminosity obtained correcting for the obscuration may be underestimated \citep{Lanzuisi2015a}.  

We have tested this hypothesis by reanalysing the \xray\ spectra of three sources showing a significant difference with respect to the 1:1 relation in $L_{bol}^{AGN}(IR)/L_{bol}^{AGN}(X)$ and having hard \xray\ (i.e., \nustar) data available.
Firstly, we present the broad-band \xray\ spectral analysis using observations taken by \nustar\ plus \xmm\ and \chandra, 
aimed at determining their obscuration and, hence, the true AGN intrinsic power. Secondly, we investigated the absorbing medium parameters by applying a proper, physically motivated torus model to \xray\ data. 

The paper is organized as follows. We present the 12MGS data set and the results from the work of \citetalias{Gruppioni2016} which led to the analysis conducted in this paper in Section~2. Section~3 describes our \xray\ broad-band data analysis and spectral fitting methods, and Section~4 gives the main results obtained in the present work. Finally, we discuss the results in a broader context and present our conclusions in Section~5.\\
Throughout this paper, we adopt a $\Lambda$CDM cosmology with a Hubble constant $H_{0}$=71 km~s$^{-1}$~Mpc$^{-1}$, $\Omega_{\textrm{m}}=0.27$ and $\Omega_{\Lambda}=0.73$.

\section{The sample}
\subsection{The local 12MGS sample of Seyfert galaxies}
The starting point of the analysis carried out by \citetalias{Gruppioni2016} was the extended 12MGS galaxy sample selected by \citeauthor{Rush1993} (\textcolor{blue}{1993}; hereafter RMS) from the \textit{Infrared Astronomical Satellite} (\iras) Faint Source Catalogue Version-2 (FSC-2). The original RMS sample comprises 893 galaxies with \iras\ 12\micron\ flux density $>0.22$~Jy; 116 of these sources were classified as AGN (51 Seyfert 1s or quasars, 63 Seyfert 2s and two blazars). This can be considered one of the largest IR-selected AGN samples with limited selection bias, likely to be almost representative of the true number and fractions of different active galaxy types.  
%
The 12\micron\ sample benefits from an extensive ancillary dataset, including photometry and spectroscopy from \xray\ to radio frequencies, coming from different observational campaigns spread over the past 20 years.

\subsection{Broad-band SED decomposition: the subsample of the 12MGS}
Thanks to the very detailed data (IR spectrum from \spitzer\ and photometry along the whole electromagnetic spectrum) available for this local sample, \citetalias{Gruppioni2016} were able to perform a detailed broad-band SED decomposition including the emission of stars, dust heated by star formation and possible AGN emission. They constrained the key physical quantities characterizing the AGN and their host galaxy, i.e., AGN luminosity and fractional contribution to the overall SED, star-formation rate (SFR) and stellar mass.

\begin{figure}
	\centering
	\includegraphics[clip=true,width=\columnwidth,trim=0.0cm 0.0cm 0.0cm 0.0cm]{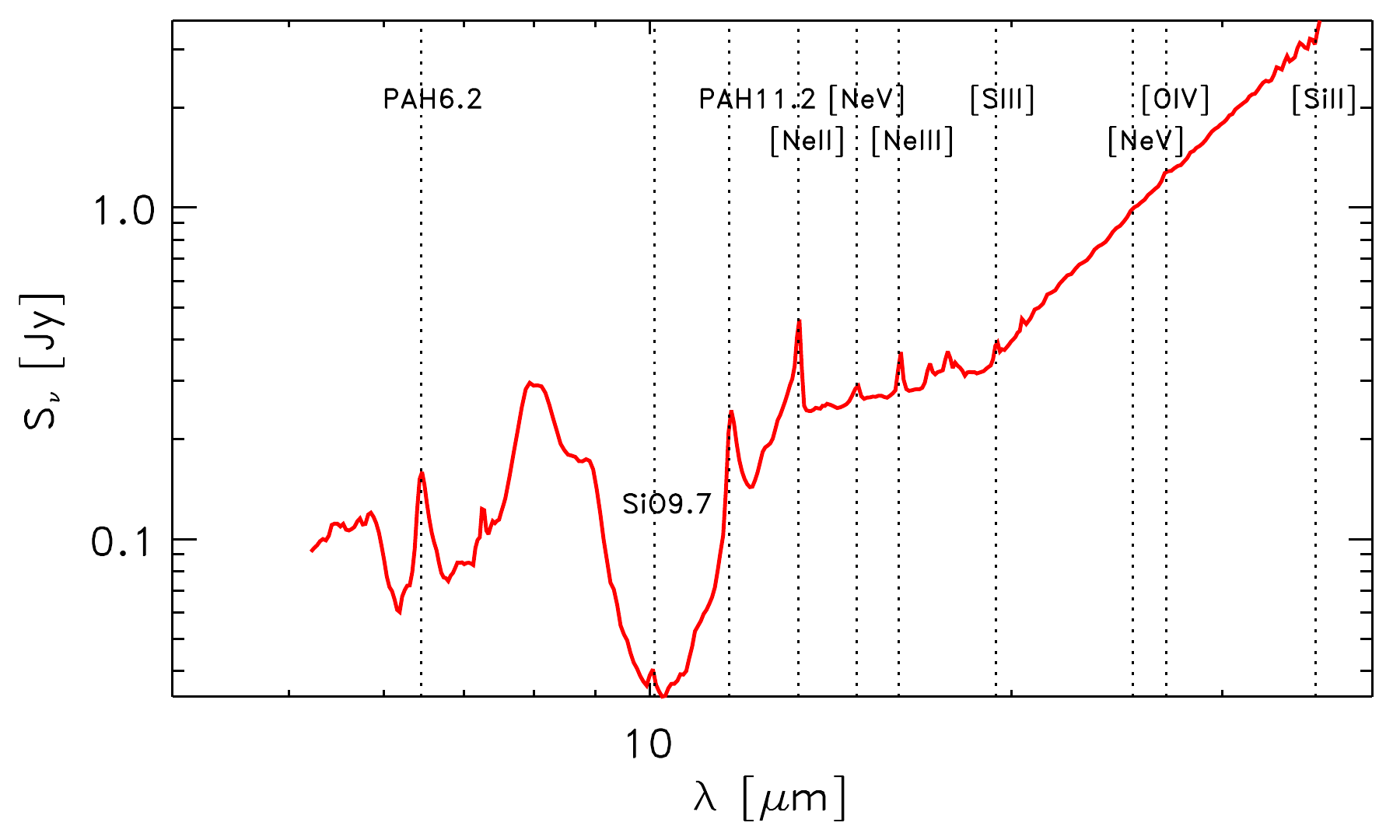}
	\includegraphics[clip=true,width=\columnwidth,trim=0.0cm 0.0cm 0.0cm 0.0cm]{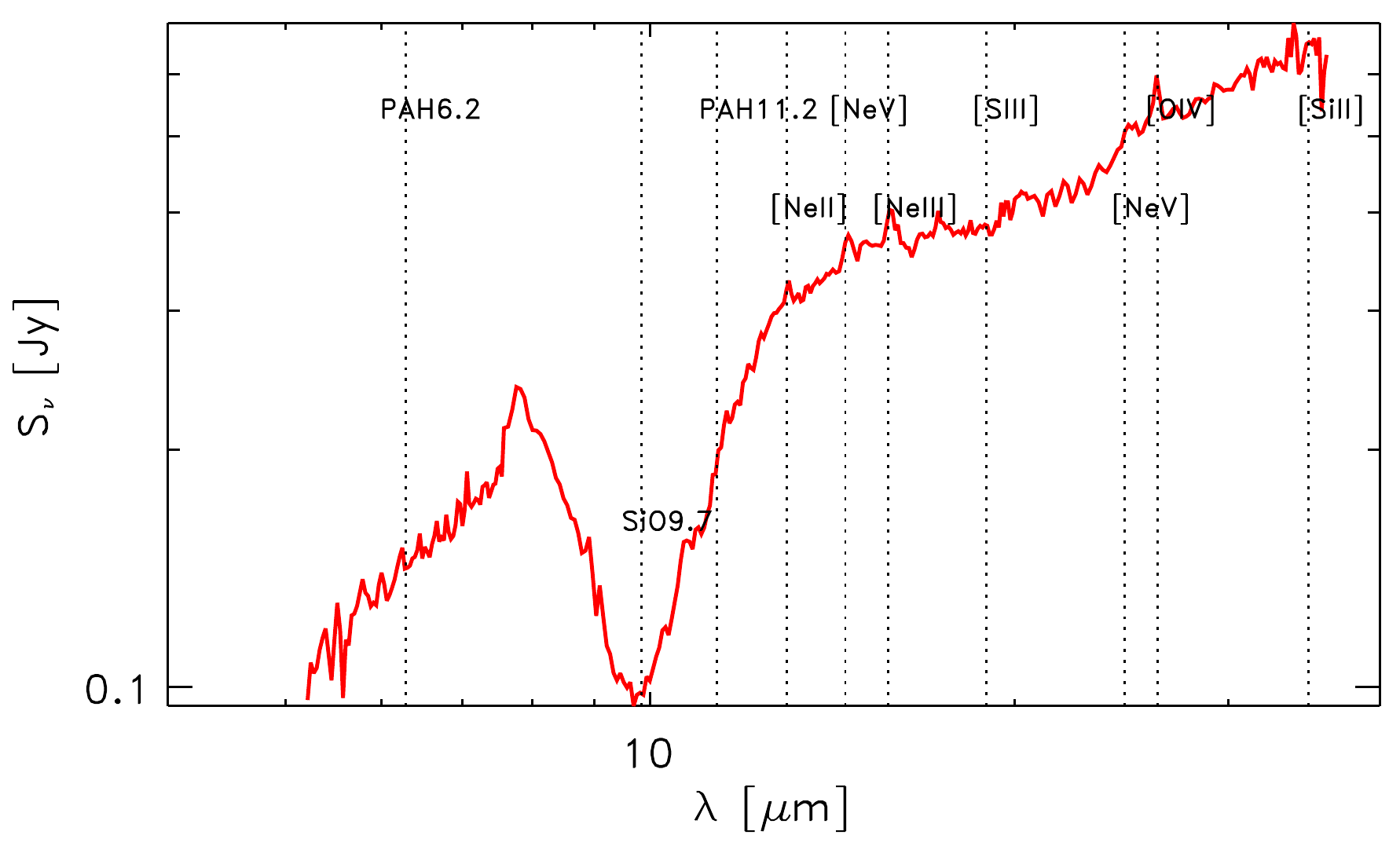}
	\includegraphics[clip=true,width=\columnwidth,trim=0.0cm 0.0cm 0.0cm 0.0cm]{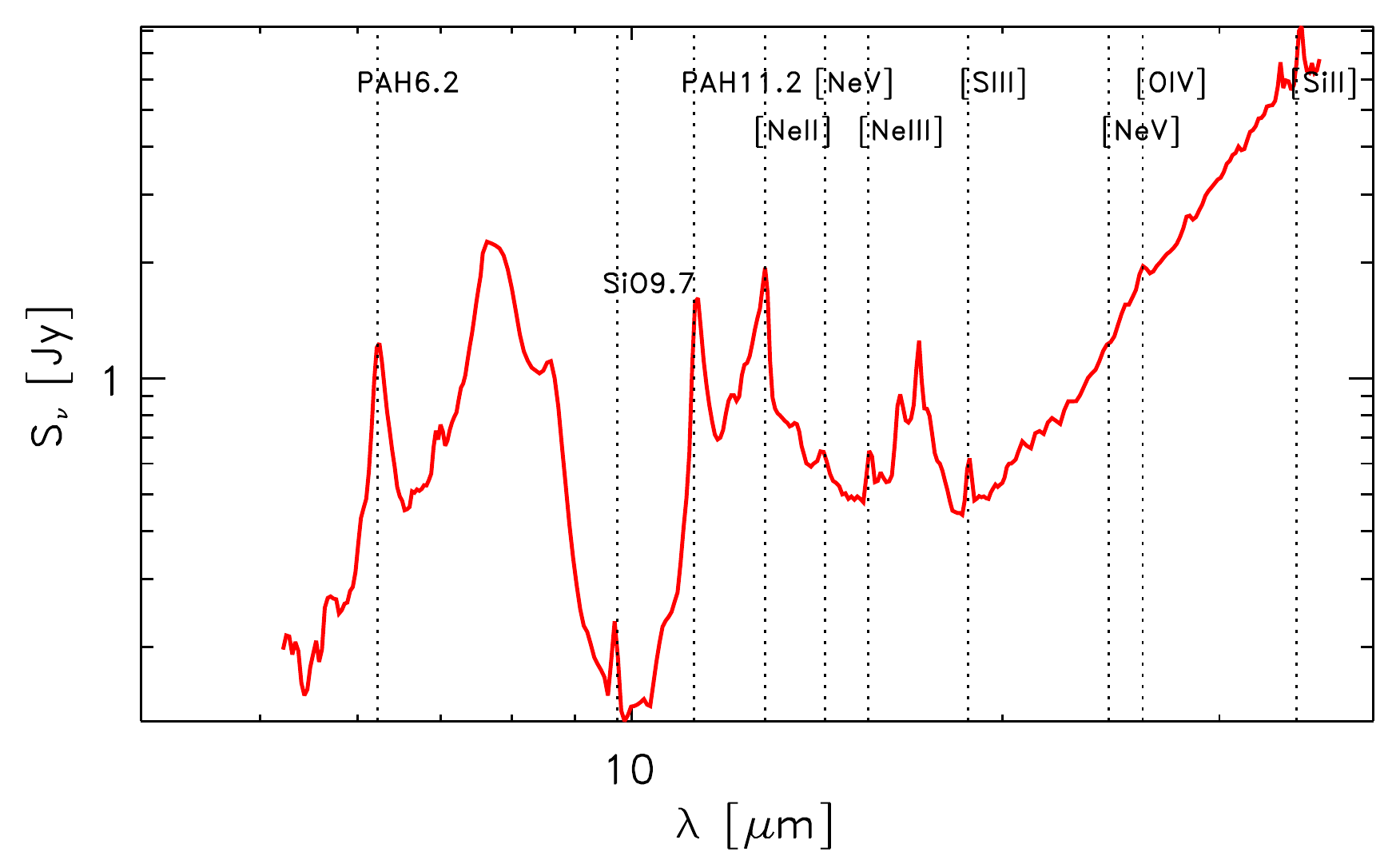}
	\vglue-0.1cm
	\caption[]{\spitzer-IRS spectra of the 12MGS sources presented in this paper (UGC05101, NGC1194 and NGC3079, from top to bottom). These are example spectra in which it is clearly visible the strong absorption feature at 9.7\micron\ associated with silicates and indicative of heavy extinction by dust. Many of the visible lines such as [Ne II], [S III] and PAH are primarily due to star formation.}
	\label{fig:IRSspectra}
\end{figure}

The availability of \spitzer-IRS low-resolution mid-IR spectra is crucial for disentangling and characterizing the AGN component (dusty torus) at its peak wavelengths and constraining its main parameters. This has been carried out by considering high optical depth ($\tau_{9.7\micron}$) templates for sources showing a strong absorption feature in their mid-IR spectra. Generally, high $N_{H}$ values are associated with a deep dust absorption feature at 9.7\micron\ attributed to silicates. Figure~\ref{fig:IRSspectra} shows the IRS spectra of the sources analysed in this paper, for which the obscured nature is suggested by the prominence of this feature seen in absorption 
at 9.7\micron\ rest frame. 

\citetalias{Gruppioni2016} obtained reliable spectra for 76 sources: 42 Seyfert~1 (including 4 quasars), 27 Seyfert~2 and 7 non-Seyfert galaxies (6 H~II and one LINER). To maximize the data coverage in the mid-IR range, where the emission from the AGN torus is expected to peak, they have re-binned the IRS spectral data in 2-\micron\ intervals and added these data to the photometric datapoints. 
About the AGN emission, 
the adopted library includes both the emission of the dusty torus and that of the accretion disk (whose contribution is expected to be more relevant in Seyfert~1 than in Seyfert~2 galaxies). The AGN models, based on a continuous (e.g., ``smooth") distribution of dust in the torus \citep{Fritz2006, Feltre2012}, were found to reproduce adequately the mid-IR emission of AGN in past works (see, e.g., \citealt{Vignali2009, Pozzi2012, Gilli2014}). To reduce the parameter degeneracies (which are present in all SED-fitting/decomposition codes), also the \xray\ classification, including the column density and \xray\ luminosity (as available from the IPAC-NED public database) has been taken into account, along with the classification based on optical and mid-IR spectra, to select the type of torus models used to fit the SED. 
In particular, for each source G16 used the $N_{H}$ value and the presence/absence of the 9.7\micron\ feature to reduce the number of possible torus solutions: they excluded all the torus models with no 9.7\micron\ absorption feature for objects with high $N_{H}$ and/or with that feature observed in the IRS spectrum. Vice versa, they excluded the 
high-$\tau_{9.7\micron}$ models for objects with low column density and/or no 9.7\micron\ feature in absorption in the IRS spectrum. AGN bolometric luminosities were estimated using the \xray\ luminosity and a bolometric correction. 

\begin{table*}
	\caption{Subsample of the 12MGS for which $L_{[2-10]\ keV}$, from \citetalias{BrightmanNandra2011a} (based on \xmm\ observations), were available. These have been used for the bolometric correction and the AGN luminosity calculations used for the relation reported in Figure~\ref{fig:Fig6Gruppioni}. Column (1): galaxy name; (2) and (3): right ascension and declination (J2000); (4): redshift; (5): logarithm of the AGN bolometric luminosity, derived from the 2--10~keV NED luminosity, used by \citetalias{Gruppioni2016}, in units of \lum; (6): 2--10~keV luminosity as reported in \citetalias{BrightmanNandra2011a} in units of \lum; (7): logarithm of the AGN bolometric luminosity obtained from the best-fitting torus model \citepalias{Gruppioni2016} in units of $L_{\odot}$; (8): logarithm of the AGN bolometric luminosity that we derived from the intrinsic 2--10~keV luminosity reported by \citetalias{BrightmanNandra2011a} after applying uniformly \citetalias{Marconi2004} bolometric correction, in units of $L_{\odot}$.}
	\label{table:12MGSsubsample}
	\begin{tabular}{lrrrcccc}
		\hline \hline
		\multicolumn{1}{c}{Name} & \multicolumn{1}{c}{RA (J2000)} & \multicolumn{1}{c}{Dec. (J2000)} & \multicolumn{1}{c}{z} &
		\multicolumn{1}{c}{$ \log L_{bol} $} & \multicolumn{1}{c}{$ \log L_{[2-10]\ keV} $} &
		\multicolumn{1}{c}{$ \log L_{bol}^{AGN}(IR) $} & \multicolumn{1}{c}{$ \log L_{bol}^{AGN}(X) $} \\
		\multicolumn{1}{c}{(1)} & \multicolumn{1}{c}{(2)} & \multicolumn{1}{c}{(3)} & \multicolumn{1}{c}{(4)} &
		\multicolumn{1}{c}{(5)} & \multicolumn{1}{c}{(6)} &
		\multicolumn{1}{c}{(7)} &
		\multicolumn{1}{c}{(8)} \\
		\hline
		3C120 & 68.294 & 5.355 & 0.0330 & 45.9 & 44.2 & 11.5 & 12.2\\
		3C273 & 187.282 & 2.051 & 0.1583 & 48.1 & 45.8 & 13.0 & 14.3\\
		3C445 & 335.955 & -2.104 & 0.0562 & 46.2 & 43.9 & 11.7 & 11.8\\
		ESO362-G018 & 79.903 & -32.658 & 0.0124 & 43.8 & 42.4 & 10.7 & 9.9\\
		IC4329A & 207.331 & -30.310 & 0.0161 & 45.5 & 43.7 & 11.5 & 11.5\\
		IRASF07599+6508 & 121.138 & 64.997 & 0.1488 & 42.9 & 42.1 & 12.4 & 9.5\\
		IRASF13349+2438 & 204.325 & 24.386 & 0.1076 & 45.5 & 43.8 & 12.7 & 11.7\\
		Izw001 & 13.396 & 12.693 & 0.0611 & 45.3 & 43.6 & 12.3 & 11.4\\
		MGC-03-34-064 & 200.598 & -16.726 & 0.0165 & 45.2 & 42.9 & 11.5 & 10.5\\
		MRK0006 & 103.051 & 74.427 & 0.0188 & 44.9 & 43.1 & 10.4 & 10.8\\
		MRK0079 & 115.637 & 49.808 & 0.0222 & 44.9 & 43.8 & 11.6 & 11.7\\
		MRK0273 & 206.174 & 55.887 & 0.0378 & 44.3 & 42.8 & 11.8 & 10.4\\
		MRK0335 & 1.576 & 20.201 & 0.0258 & 44.7 & 43.5 & 10.9 & 11.3\\
		MRK0463 & 209.011 & 18.372 & 0.0504 & 44.8 & 43.1 & 12.5 & 10.8\\
		MRK0704 & 139.607 & 16.307 & 0.0292 & 44.5 & 43.4 & 11.3 & 11.1\\
		NGC0262 & 12.201 & 31.957 & 0.0150 & 44.2 & 43.3 & 11.3 & 11.0\\
		NGC0424 & 17.862 & -38.085 & 0.0118 & 44.3 & 42.5 & 11.2 & 10.0\\
		NGC0526A & 20.978 & -35.065 & 0.0191 & 44.8 & 43.3 & 10.3 & 11.0\\
		NGC1194 & 45.952 & -1.104 & 0.0136 & 43.9 & 42.3 & 11.1 & 9.7\\
		NGC1365 & 53.402 & -36.140 & 0.0055 & 43.0 & 42.5 & 10.4 & 10.0\\
		NGC2992 & 146.418 & -14.323 & 0.0077 & 44.0 & 43.1 & 10.1 & 10.8\\
		NGC3079 & 150.491 & 55.681 & 0.0037 & 41.2 & 40.9 & 10.0 & 8.2\\
		NGC3516 & 166.703 & 72.567 & 0.0088 & 44.3 & 43.8 & 10.2 & 11.7\\
		NGC4051 & 180.791 & 44.531 & 0.0023 & 42.6 & 40.9 & 8.9 & 8.2\\
		NGC4151 & 182.644 & 39.402 & 0.0033 & 44.0 & 42.1 & 10.5 & 9.5\\
		NGC4253 & 184.608 & 29.814 & 0.0129 & 44.3 & 43.0 & 10.6 & 10.6\\
		NGC4388 & 186.441 & 12.664 & 0.0084 & 43.3 & 42.9 & 10.7 & 10.5\\
		NGC4593 & 189.917 & -5.346 & 0.0090 & 44.2 & 42.8 & 10.0 & 10.4\\
		NGC5256 & 204.576 & 48.275 & 0.0279 & 43.2 & 42.2 & 11.2 & 9.7\\
		NGC5506 & 213.309 & -3.207 & 0.0062 & 44.4 & 42.8 & 10.6 & 10.4\\
		NGC5548 & 214.499 & 25.137 & 0.0172 & 45.1 & 43.4 & 11.3 & 11.1\\
		NGC6810 & 295.890 & -58.656 & 0.0068 & 40.6 & 39.9 & 10.0 & 7.1\\
		NGC6890 & 304.577 & -44.806 & 0.0081 & 41.7 & 42.2 & 9.7 & 9.7\\
		NGC7213 & 332.319 & -47.166 & 0.0058 & 43.4 & 42.7 & 9.7 & 10.3\\
		NGC7469 & 345.815 & 8.874 & 0.0163 & 44.5 & 43.2 & 10.9 & 10.9\\
		UGC05101 & 143.953 & 61.356 & 0.0394 & 42.3 & 42.5 & 12.0 & 10.0\\
		\hline
	\end{tabular}
\end{table*}

\subsubsection{AGN bolometric luminosity: $L_{bol}^{AGN} (IR) - L_{bol}^{AGN} (X)$ relation and source selection}

\begin{figure}
	\centering
    \includegraphics[width=\columnwidth]{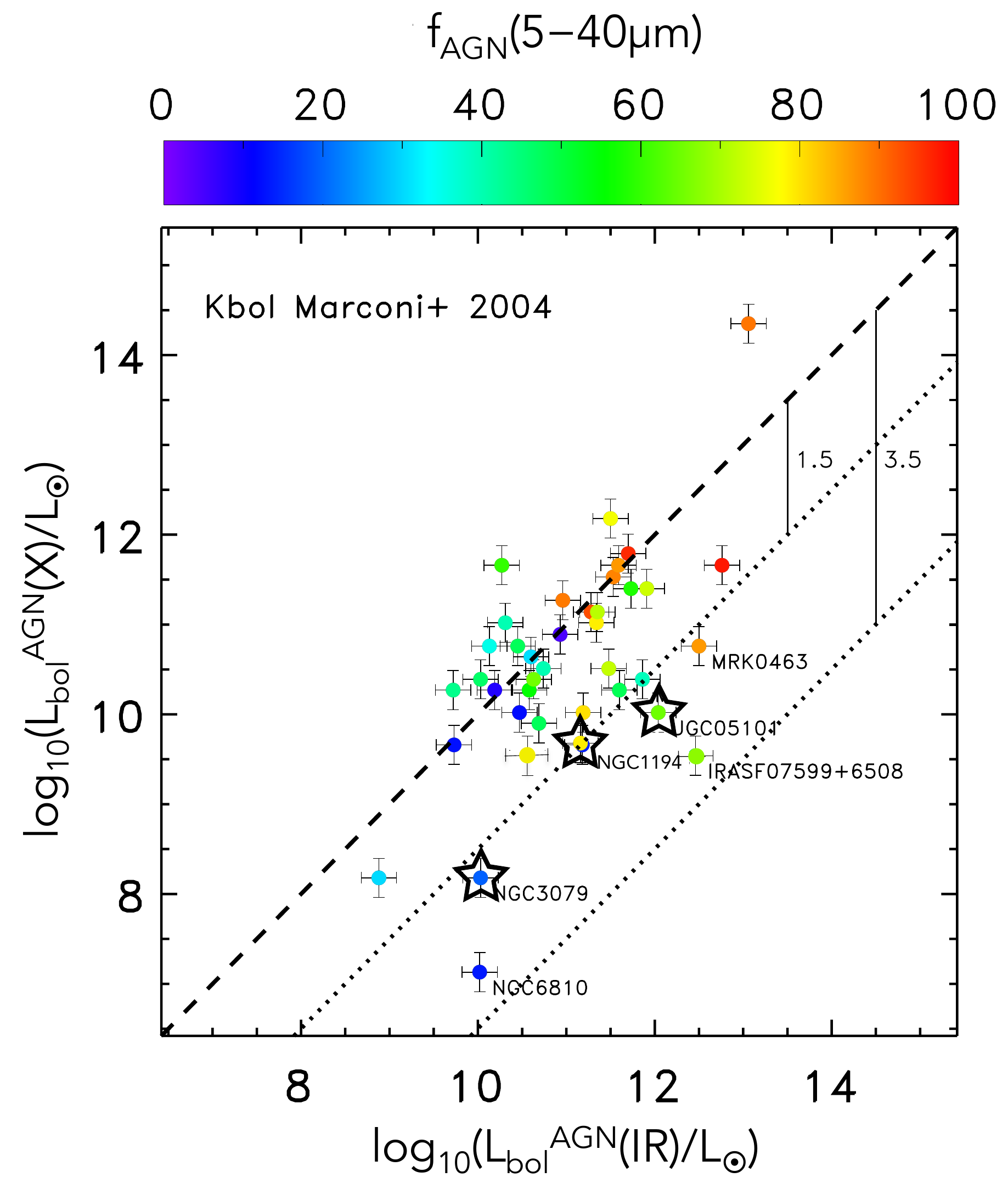}
	\vglue-0.15cm
	\caption{AGN bolometric luminosity obtained from the best-fitting torus model \citepalias{Gruppioni2016} versus AGN bolometric luminosity that we have computed from the 2--10~keV luminosity reported by \citetalias{BrightmanNandra2011a} using the \citetalias{Marconi2004} bolometric correction. The considered sources are listed in Table~\ref{table:12MGSsubsample}. The symbols shown in color-gradient represent sources \citepalias[from the 76 galaxy sample of][]{Gruppioni2016} with different fractions of the luminosity produced by the AGN in the rest-frame 5--40\micron\ wavelength range, as shown in the top. The dashed line represents the 1:1 relation; the two dotted lines show the $L_{bol}^{AGN}(X)$ scaled down by a factor 1.5--3.5 in log scale, as indicated by the vertical bars. The most extreme sources are labeled, 
and the three targets of this work are encircled with empty stars.}
	\label{fig:Fig6Gruppioni}
\end{figure}

In Fig.~6 of \citetalias{Gruppioni2016} the AGN bolometric luminosity derived from the SED decomposition (i.e., from the best-fitting torus model) is compared to the same quantity obtained from the 2--10~keV luminosity (IPAC-NED database). This relation is more dispersed and with a lower significance than those derived using mid-IR AGN lines in the \citetalias{Gruppioni2016} work. 
The scatter around the 1:1 relations is $\sigma_{X}=1.1$, $\sigma_{[Ne V]14.3}=0.61$, $\sigma_{[Ne V]24.3}=0.66$, and $\sigma_{[O IV]25.9}=0.70$. 
The significance of these relations are 1.5$\sigma$, 3.4$\sigma$, 3.3$\sigma$,  and 3.2$\sigma$ for X-ray, [Ne V]14.3\micron, [Ne V]24.3\micron\ and [O IV]25.9\micron\ luminosities, respectively. The relative low significance of the relation involving the \xray\ luminosity is ascribed to the presence of several objects showing AGN bolometric luminosities estimated from the \xray\ band (by means of a bolometric correction) lower than those derived by decomposing the SED. Most of the deviating objects are known to be highly obscured AGN (e.g., IRASF07599+6508, UGC05101), showing relatively low \xray\ luminosities ($<10^{43.5}$~\lum). However, at these \xray\ luminosities it is possible that some of these sources may present a significant contamination from star-formation (e.g., \citealt{BrightmanNandra2011b} consider ``\xray\ AGN" only sources with an observed 2--10~keV luminosity larger than $10^{42}$~\lum). 

\label{sec:selection_oursubsample}
As the adopted \xray\ luminosities have been corrected for absorption, $L_{X}$ should represent the intrinsic emission from the nucleus, and the $L_{bol}^{AGN}(IR)/L_{bol}^{AGN}(X)$ ratios should be close to 1. The difference up to three orders of magnitude shown in the plot for some sources suggests that correcting for obscuration may be difficult also in X-rays if the data quality is not high and the observing band is limited (i.e., data were available only below 10~keV). The intrinsic \xray\ luminosity obtained by means of a correction for obscuration could be underestimated.

Since the intrinsic luminosity in the 2--10~keV band has been taken from various literature papers, we tried to evaluate how much 
of the scatter found in \citetalias{Gruppioni2016} could result from a not uniform choice of the $k_{bol}$. 
We computed the bolometric correction as a function of the bolometric luminosity considering the 2--10~keV luminosity. The AGN sample is composed of sources for which the $L_{[2-10]\ keV}$ values, taken from \citeauthor{BrightmanNandra2011a} (\textcolor{blue}{2011a}, hereafter BN11a) and based on \xmm\ observations, were available. Table~\ref{table:12MGSsubsample} provides the relevant information about these sources. Despite the scatter, the known trend of increasing bolometric correction at increasing bolometric luminosity is confirmed (\citeauthor{Lusso2012};  
see also the recent work by \citealt{Martocchia2017}). 
Since it seems that for not all of the sources the same $k_{bol}$ was applied, to uniform the analysis we systematically re-computed all the bolometric luminosities by adopting \citetalias{Marconi2004} relation; Table~\ref{table:12MGSsubsample}, column 8, lists the derived values.

Now we can plot the $L_{bol}^{AGN}(IR)$ versus $ L_{bol}^{AGN}(X)$ using in the y-axis our newly X-ray-derived bolometric values (Fig. \ref{fig:Fig6Gruppioni}). The dashed line on this diagram represents the 1:1 relation. The correlation is tighter with respect to the original one; nonetheless, there are still a few sources clearly ``underluminous" in the 2--10~keV band. These galaxies are labeled in the plot. 

We focused on the sources showing the most striking separation from the 1:1 relation ($\sim1.5-3.5$ times below in the log--log space). Among these, we selected those for which observations from at least two satellites were available, one of these being \nustar\ \citep{Harrison2013}, complementary to \chandra\ and/or \xmm, to have a broad-band, although not simultaneous, \xray\ spectral coverage. The resulting selected sources are 
UGC05101, NGC1194 and NGC3079. We note that NGC3079 was already reported as a Compton-thick AGN by \citetalias{BrightmanNandra2011a} in virtue of its large iron K$\alpha$ EW and reflection fraction. 
Furthermore, NGC1194 and NGC3079 are water megamaser AGN \citep{Pesce2015, Kondratko2005} and, as  such, were presented in a compilation by \citeauthor{Masini2016} \textcolor{blue}{2016} (hereafter M16). For instance, the almost ubiquitous presence of heavy obscuration in water megamasers AGN was presented in several works (e.g., \citealt{Greenhill2008, Castangia2013, Leiter2018}). 
For UGC05101 and NGC3079, data from all three satellites are available, while NGC1194 has \xmm\ and \nustar\ observations only. 

We note that there are other sources apparently satisfying the adopted conditions for ``extreme" source selection. These either lack \nustar\ data or are undetected by \nustar; in the case of Mrk463, a comprehensive, broad-band analysis of the galaxy hosting two active nuclei \citep{Bianchi2008} has been recently published \citep{Yamada2018}. 
Focusing on the two sources deviating most from the 1:1 relation, i.e. NGC6810 and IRASF07599+6508, the former is most likely a starburst according to published \xray\ observations \citep[][\citetalias{BrightmanNandra2011a}]{Strickland2007} and has no available \nustar\ data, while the latter is a Broad Absorption Line Quasi-Stellar Object (BALQSO), undetected in the 28~ks \nustar\ observation \citep{Luo2014}.

\section{\xray\ broad-band spectral analysis}

\subsection{Data reduction}
The set of observations used in this work consists of European Photon Imaging Camera (EPIC-pn and MOS) \xmm\ observations along with the \chandra\ Advanced CCD Imaging Spectrometer (ACIS-S) and the \nustar\ Focal Plane Modules (FPMA-B) observations.

\subsubsection{XMM-Newton}
The raw observation data files have been processed using SAS\footnote{https://www.cosmos.esa.int/web/xmm-newton/sas.} 
v16.0 tasks, with the associated latest calibration files. A standard \xmm\ reduction has been applied and the calibrated event files were screened to check for soft protons flaring examining the light curves at energy $>$10~keV. For NGC3079 the cleaning was done applying a threshold to the count rate (CR$<$1.5 for pn and CR$<$0.7 for the MOS cameras). The source spectra were extracted from a circular region centered on each source (radius of 20\arcsec\ and 10\arcsec\ for pn and MOS events, respectively, for NGC1194; 30\arcsec\ and 20\arcsec\ for pn and MOS events for NGC3079; 20\arcsec\ for both pn and MOS events in the case of UGC05101). The chosen areas encircle $\sim$80\% of the photons energy at 
5.0~keV. For the background we typically chose a 10 times larger area (when possible), adopting circular and elliptical regions on the same CCD where the source is located, ensuring no contamination from nearby sources. This guarantees a good sampling of the background spectrum and optimize the signal-to-noise ratio. None of the sources shows photon pile-up. Here and in the following \xray\ spectral fits, the background is obviously normalized to the source area. 

\subsubsection{Chandra}
To perform the data processing and calibration we used the analysis package CIAO\footnote{http://cxc.harvard.edu/ciao.} v4.8. All the observations were carried out in imaging mode with the ACIS-S3 detector operating in faint mode. Starting from level 1 event files, new level 2 filtered event files were generated, using the latest calibrations. From the evt2 files we generated new files selecting good grades, the CCD in use and data in the $\sim0.5-7$~keV spectral range. \chandra\ images also show diffuse emission from the host galaxy, unresolved by \xmm, for their superior spatial resolution. 
In particular, the \chandra\ image of NGC3079 shows that the nuclear source is embedded in a bubble of diffuse emission \citep[stellar winds and supernova explosions,][]{Cappi2006}. This structure extends for $\sim$11\arcsec\ around the nuclear position; however, it does not contribute significantly to the nuclear emission at E$>$2~keV. Therefore, we extracted the source spectrum from a circular region of radius 4\arcsec\ centered on the source maximizing the nuclear emission. For UGC05101 we varied the extraction radius, obtaining \chandra\ spectra from two circular regions, one including mostly the nuclear emission (radius of 2\arcsec) and the other encircling also a more extended emission (radius of 5\arcsec).  For the background, we chose an appropriate large area in both cases, to be free from contaminating sources, always following the same criteria. We then ran {\sc specextract} tool to create spectra and responses matrices.

\subsubsection{NuSTAR}
The data have been reduced using {\sc NUSTARDAS}\footnote{https://heasarc.gsfc.nasa.gov/docs/nustar/analysis.} v1.8 and instrumental responses from \nustar\ CALDB. We produced calibrated and cleaned event files from the level 1 event files applying the {\sc nupipeline} script that runs in sequence all the tasks of the standard data processing. Events are flagged according to various cuts criteria and screened by applying cleaning criteria. We produced source extraction region files by choosing a circular region of 40, 45 and 30 arcsec radius centered on the source for NGC1194, NGC3079 and UGC05101, respectively, to optimize the signal-to-noise ratio (SNR). The same choice was made for FPMA and FPMB. We selected the background regions in blank areas as close to each source as possible without including source photons, on the same detector where the source is located and away from the outer edges of the field of view (which have systematically higher background). To extract the level 3 science products, e.g. spectra and response matrices, we run the {\sc nuproducts} module.

\subsection{Photon statistics and spectral binning}
For each source observation, the incidence of the source counts with respect to the total (source$+$background) is $>$80\% in the spectral range used to fit the data. The spectra of the two individual MOS cameras were summed into a single spectrum for each observation, summing also the corresponding background spectra, and merging the response matrices as well; response matrices are weighted by the exposure time of each instrument. We will fit the summed MOS spectra simultaneously with the pn spectrum. This gives the same quantitative results as fitting the three EPIC spectra simultaneously,  and the SNR in each bin increases. Data from the two \nustar\ FPM detectors will be fitted simultaneously rather than coadded because data are not always completely consistent with each other over the whole range in which they operate.

Finally, we rebinned the spectra for further analysis with {\sc XSPEC} (version 12.9.0; \citealt{Arnaud1996}), in order to have at least 20 (15 in case of \chandra\ data) counts in  each spectral bin. The binning we applied is appropriate for $\chi^{2}$ statistics that is valid in Gaussian statistics regime, e.g., if there are enough counts in each spectral bin.

\subsection{Spectral modeling}
\label{sec:Spectral_modeling}
In order to derive the source physical properties, within the limits of the available statistics, we first constructed a phenomenological model to obtain a first-order source characterization.
We then adopted a physically motivated model, {\sc MYTorus} \citep{Murphy2009}, specifically developed to model Compton-thick obscuration 
in a toroidal geometry. We developed a systematic approach to the spectral fitting of our sample, starting the analysis from the \xmm\ pn data to have preliminary spectral indications of the source properties using the dataset with the highest statistics; these indications were then verified including also the MOS data in the analysis. 
Then, separately, the same best-fitting model obtained with \xmm\ has been applied to \chandra\ and \nustar\ data separately to verify whether  
significant spectral variations are present among the various datasets (having in mind the different spectral coverage and energy resolution of the instruments). Once checked that the same model applies reasonably well to all datasets, as a final stage we performed a full-band spectral fit considering all the available \xray\ data. 
The three sources seem to be characterized by similar spectral components, therefore the adopted spectral modeling will be the same. This approach, testing increasingly complicated models to find the best spectral fit to the data, is described below. 

As a phenomenological model, we initially assumed a standard power-law model, to account for the source continuum emission, and Galactic absorption, fixed to the measured value obtained using {\sc nh} tool within FTOOLS \citep{Kalberla2005}.  
This is the simplest scenario for an AGN. We interpret any relatively flat \xray\ slope (power-law photon index $\Gamma$ with values $\lesssim$1) as suggestive of the presence of either a reflection or a transmission component (i.e., passed through an absorbing medium), or both. 
In the following phenomenological model we will assume the transmission scenario (to limit model degeneracies and make the model not too complicated at this first stage), although we note that reflection can be present (as in the case of NGC1194). We then gradually included additional, physically motivated model components looking at the residuals (data-model in units of the statistical uncertainty, reported in the lower panel of each figure showing the \xray\ spectra and data). 
For what concerns the soft \xray\ band, ``extra" \xray\ emission (wrt. the extrapolation of the primary power-law emission) is generally observed and interpreted as nuclear emission scattered at the ``poles" by electrons (Thomson scattering) plus  thermal emission ascribed to the host galaxy. Finally, once the continuum is fitted relatively well, we tried to model any residuals indicating the possible presence of emission features. 
To summarize, the spectral model components used to provide a phenomenological spectral characterization of each source are the following: 

\begin{enumerate}
	\item Neutral absorption at redshift 0 (\textsc{phabs} model within {\sc xspec}), associated with the Galaxy (from \citealt{Kalberla2005}). 
 \item A neutral absorber at the source redshift $N_{H,1}$ due to absorbing structures in the host galaxy, such as dust lanes \citep[see e.g.,][]{Severgnini2015}; values are usually $\leq$few$\times10^{21}~$~cm$^{-2}$.
	\item A thermal plasma component, \textsc{mekal}, to account for the excess with respect to the power law, ubiquitously observed in the soft spectra and visible at energies $<$2~keV. It may arise from unresolved emission structures likely due to binaries and supernovae remnants.
	\item A primary power law, \textsc{powerlaw}, to account for the source continuum emission. During the spectral fitting procedure, the spectral index of the primary power law was fixed to the AGN typical value \citep[$ \Gamma_{1}=1.8 $, e.g.][]{Piconcelli2005}.
	\item A second neutral absorption component at the source redshift $N_{H,2}$, \textsc{zphabs}, for nuclear absorption intrinsic to the source. We left it as a free parameter during the model fitting procedure.
	\item A soft power law to account for soft emission from any scattered power law radiation often seen in type~2 Seyfert spectra \citep[e.g.,][]{Turner1997}. The power law index $\Gamma_{2}$ is fixed to $\Gamma_{1}$, being the photons physical origin the same. The normalization of the soft power law is typically a few percent of the primary one \citep[e.g.,][]{Lanzuisi2015b}. 
	\item Line emission modeled by one or more Gaussian components (\textsc{zgauss}). 
\end{enumerate}
We have chosen the best-fit model as the simplest one that has a combination of components with the lowest reduced $\chi^{2}$ and is physically motivated.

Given that an obscuring model seemed to be required, we subsequently performed the analysis adopting the spectral-fitting suite {\sc MYTorus} trying to reproduce in more physical terms the emission of the obscured objects. {\sc MYTorus} \citep{Murphy2009} is a physical model, designed specifically for modeling the \xray\ spectra of active galaxies assuming a toroidal geometry for the reprocessor (uniform and cold).  
{\sc MYTorus}, based on Monte Carlo simulations, self-consistently includes reflection and transmission (see e.g. \citealt{Lanzuisi2015b} for further details, Appendix~B of \citealt{Lanzuisi2015a} for a comparison with other models applied to heavily obscured AGN, and the recent work by \citealt{Marchesi2019}). 
Additional components (low-energy scattering, host galaxy emission, and emission lines, as observed in particular in the \xray\ spectrum of NGC1194) were included along with {\sc MYTorus}. 

The best-fitting models are reported source by source. All errors are quoted at the 90\% confidence level for one parameter of interest \citep[$\Delta\chi^{2}$ = 2.706,][]{Avni1976}.

\begin{figure*}
\centering
\includegraphics[width=\textwidth]{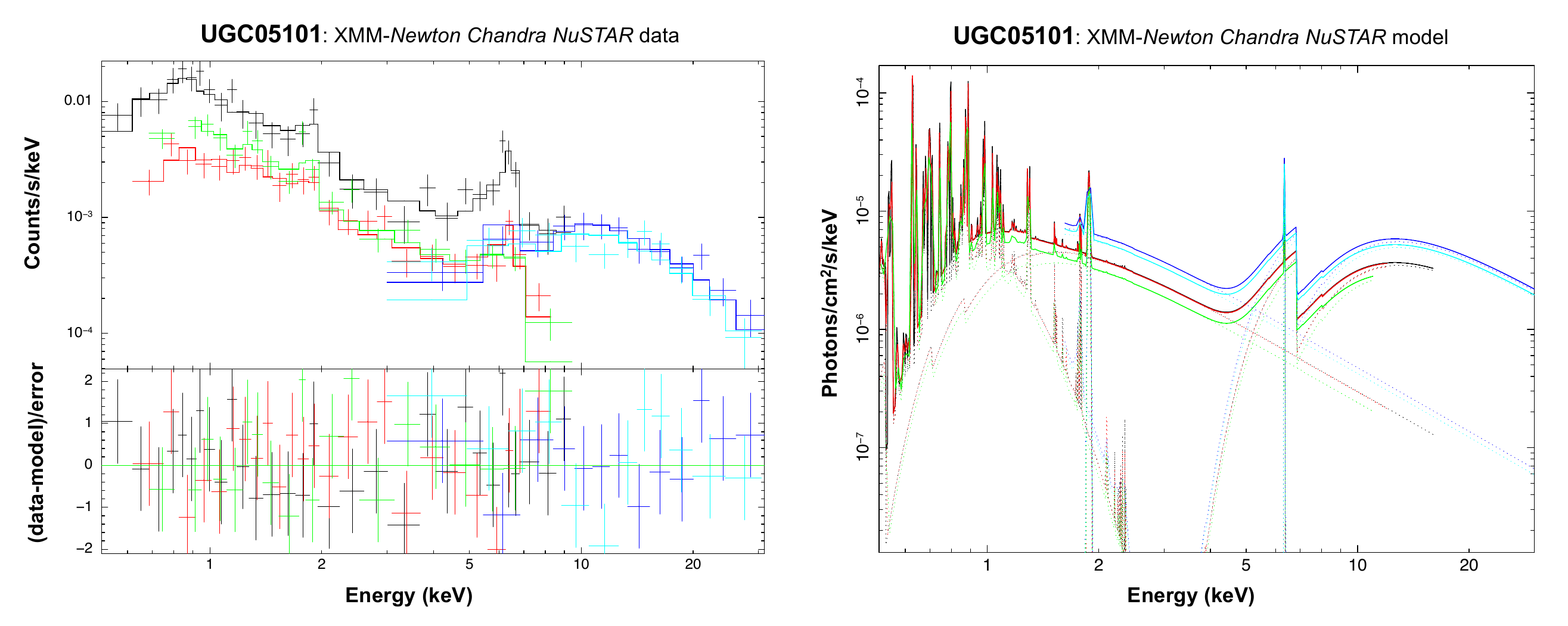}
\includegraphics[width=\textwidth]{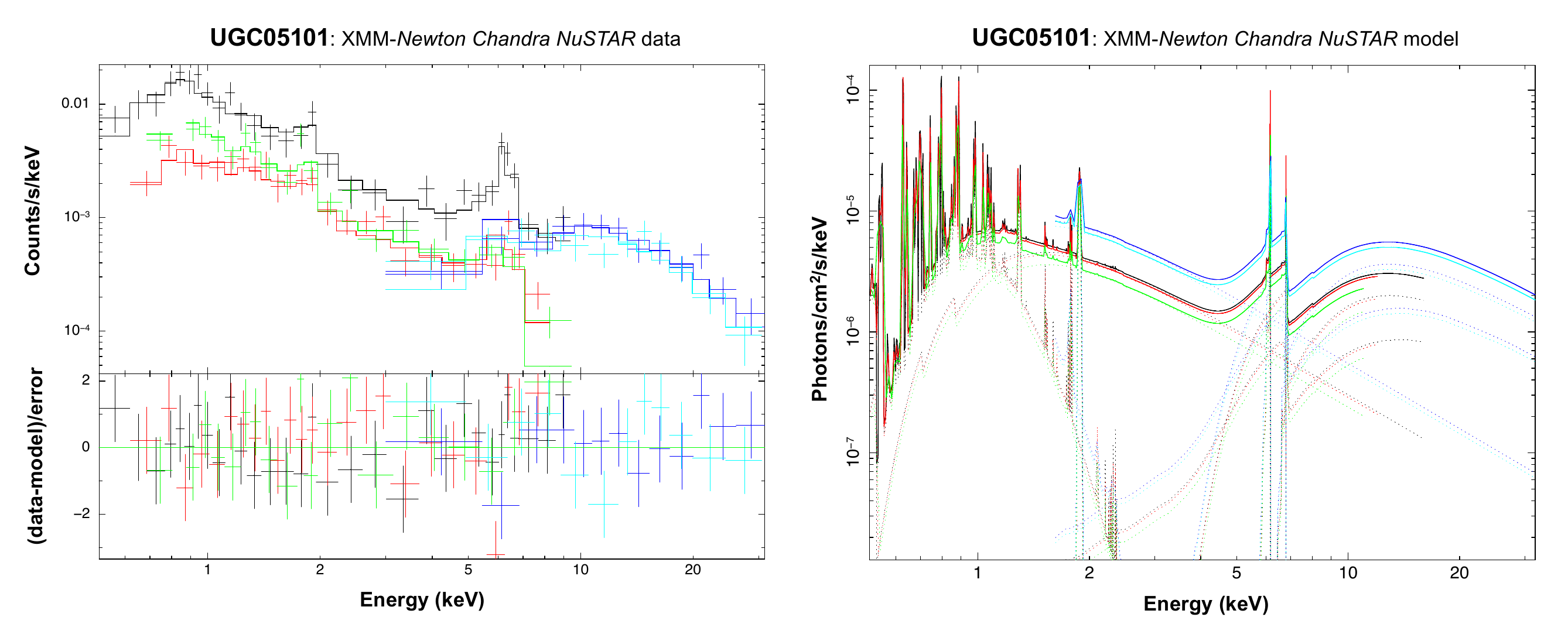}
\vglue-0.2cm
\caption[]{
{\it (Left)} 0.5--30~keV spectrum of UGC05101 (\xmm\ EPIC pn: black, EPIC MOS1$+$2: red; \chandra: green; \nustar\ FPMA and FPMB: blue and light blue, respectively), fitted with the phenomenological model (top left: {\sc phabs[mekal + zgauss + zpowerlaw + zphabs(zpowerlaw + zgauss)]}) and with {\sc MYTorus} (bottom left: {\sc phabs(mekal + zgauss + zpowerlaw + {\sc MYTorus})}), and data-model residuals in units of $\sigma$ (bottom panels of both spectra). 
{\it (Right)} Corresponding spectral-fitting models with the same color code.
}
\label{fig:UGC05101_spectra}
\end{figure*}

\subsubsection{UGC05101 spectral fits}
A first individual analysis of the spectra from each instrument shows that there is no evidence for significant spectral differences, therefore we used the datasets together, adopting at first the phenomenological model described in Sec.~3.3 and then a more physical one, as previously described. Using the simultaneous spectral fit we are able to recover information from the broad $\sim$0.5--30~keV band. In order to take into account calibration issues among the different datasets, besides some flux differences due to the fact that the analyzed data were not simultaneous, each dataset has been multiplied by a constant. The resulting values for the constants are typically consistent to within 20\%. 
A similar approach will be adopted for the other sources. 
For what concerns the {\sc MYTorus} model, it is adopted in the coupled version  (cross-normalization fixed to 1 between primary power law, reflection and line emission). 

\vglue0.2cm
\pn\textbf{Phenomenological model.} Figure~\ref{fig:UGC05101_spectra} (top-left panel) shows the full-band spectrum fitted with the phenomenological transmission model (reported in the top-right panel), including data-model residuals in units of $\sigma$. The black curve refers to the pn, while the red one shows the summed MOS1-2 spectrum; the green line indicates the ACIS-S3 spectrum, and blue and light blue lines indicate the FPMA and FPMB spectra, respectively. 

The continuum spectral components model the cut-off due to Galactic ($N_{H}$ (Gal)$=3\times10^{20}$~cm$^{-2}$) and host galaxy absorption ($N_{H,1} \sim0.8\times10^{22}$~cm$^{-2}$), plus a thermal emission with temperature kT$\sim$0.2~keV, a scattered component and a heavily absorbed primary component. 
The normalization of the soft power law ($\Gamma$ fixed to 1.8) is 3\% of the primary one at 1~keV. The absorption intrinsic to the source has a value $N_{H,2}=1.21^{+0.16}_{-0.14} \times 10^{24}$~cm$^{-2}$. Narrow Gaussian lines (\textsc{zgauss}, with $\sigma$ frozen to 10~eV) are added with best-fitting centroid energies at 1.98~keV and 6.4~keV (rest-frame energy) to model emission lines from Si~XIV and neutral Fe~K$\alpha$, respectively. 
In {\sc xspec} notation, the resulting model is {\sc phabs[mekal + zgauss + zpowerlaw + zphabs(zpowerlaw + zgauss)]}. 
In the \nustar\ spectrum, because of poor spectral resolution ($\sim400$~eV at 6~keV), the Fe emission line is not constrained. 
Apparently, the iron line energy is not consistent with being neutral; however, 
once only \xmm\ data (having good energy resolution and photon statistics at the energy of the line) are used, we find it to be in the range 6.4--6.8~keV (at the 90\% confidence level), i.e., consistent with neutral emission up to the He-like transition; we note that neutral or mildly ionized iron emission (E=6.46$\pm{0.04}$~keV) was found by \cite{Oda2017} analysis and reported in previous investigation based on \chandra\ data alone \citep{Imanishi2003, Iwasawa2011}. 
The reduced $\chi^{2}$ ($\chi^{2}$/dof, where dof=degrees of freedom) of our best fit is 80.8/87. 

\vglue0.2cm
\pn\textbf{MYTorus model.} Figure~\ref{fig:UGC05101_spectra} (bottom-left panel) shows the full-band spectrum fitted with the {\sc MYTorus} model and data-model residuals in units of $\sigma$; in {\sc xspec} notation, this model can be reported as {\sc phabs(mekal + zgauss + zpowerlaw + {\sc MYTorus})}; see Fig~\ref{fig:UGC05101_spectra} (bottom-right panel). 
The power law spectral index was fixed to the AGN typical value $\Gamma=1.8$. Table~\ref{tab:allSources} lists the basic parameter values with their relative errors. The derived 
$N_{H,2}$ value is $1.21^{+0.19}_{-0.17}\times10^{24}$~cm$^{-2}$, consistent with the recent analysis carried out by \cite{Oda2017}. 
The absorption column density along the line of sight is basically the equatorial one,
since the incident angle (angle between the observer line-of-sight and the symmetry axis of the torus) is $\theta_{obs}=85^{\circ}$.  
The reduced $\chi^{2}$ of the fit is 92.9/89. 

A more in-depth comparison of our results with those of  \cite{Oda2017}, where \chandra, \xmm, \nustar, \suzaku\ and \swift/BAT data were used and interpreted using both a  phenomenological and a more ``physical" modeling, indicates that overall our results are in agreement with theirs. More specifically, \cite{Oda2017} used the \cite{Ikeda2009} torus model, and found that its half-opening angle is $>41^{\circ}$ (see their Table~5) and the equatorial column density is $1.32^{+0.32}_{-0.37}\times10^{24}$~cm$^{-2}$. Due to their broader band and higher statistics data, they were able to constrain the photon index (1.63$^{+0.27}_{-0.15}$). Conversely, we were not able to obtain good constraints on the photon index once it was left free to vary within the {\sc MYTorus} modeling; however, $\Gamma\sim1.6$ still provides a good representation of our data, and the derived main source parameters adopting this photon index are basically unchanged within errors. 

The AGN flux in the 2--10~keV band is $\sim1.8\times10^{-13}$~\cgs\ ($\sim2\times10^{-13}$~\cgs\ in \citealt{Oda2017}) and corresponds to an intrinsic (i.e., corrected for the measured obscuration) rest-frame 2--10~keV (10--40~keV) luminosity of $\sim1.2\times10^{43}$~\lum\ ($\sim2.5\times10^{43}$~\lum); see Table~\ref{tab:allSources_FLNH}. 

\begin{figure*}
\centering
\includegraphics[width=\textwidth]{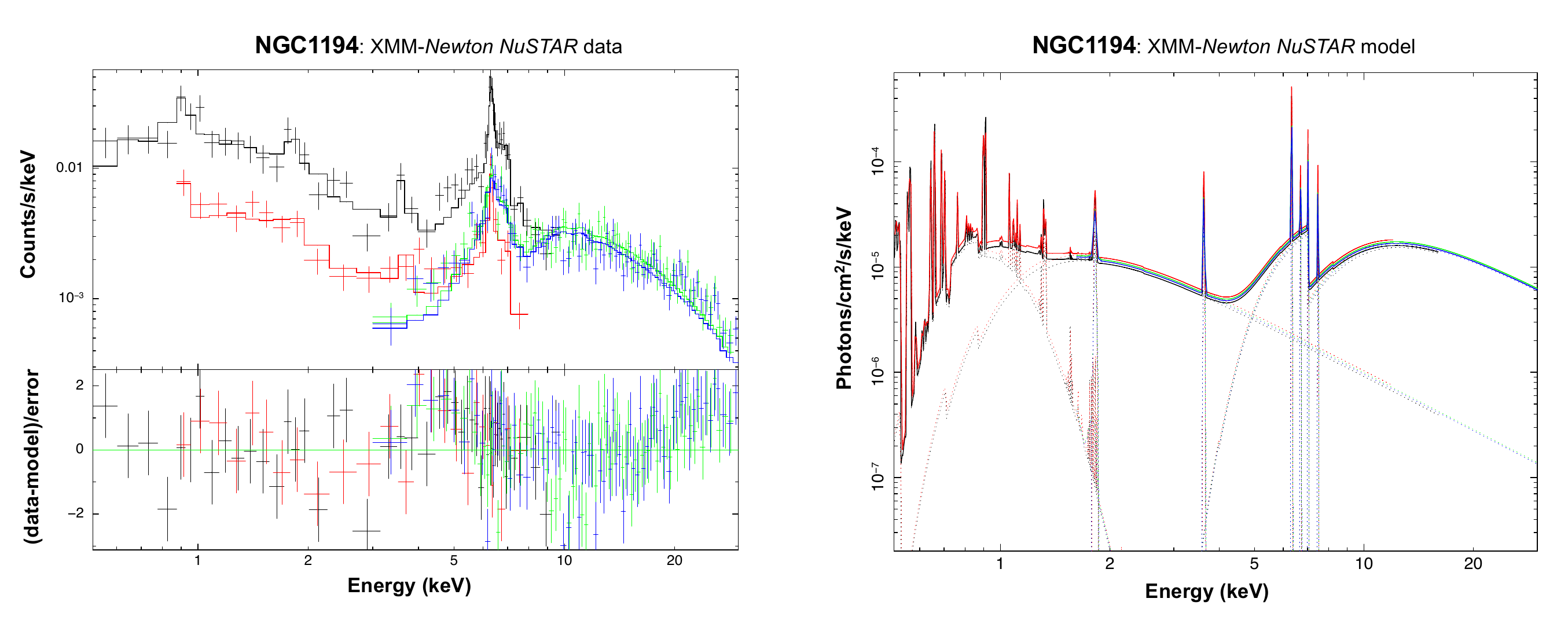}
\includegraphics[width=\textwidth]{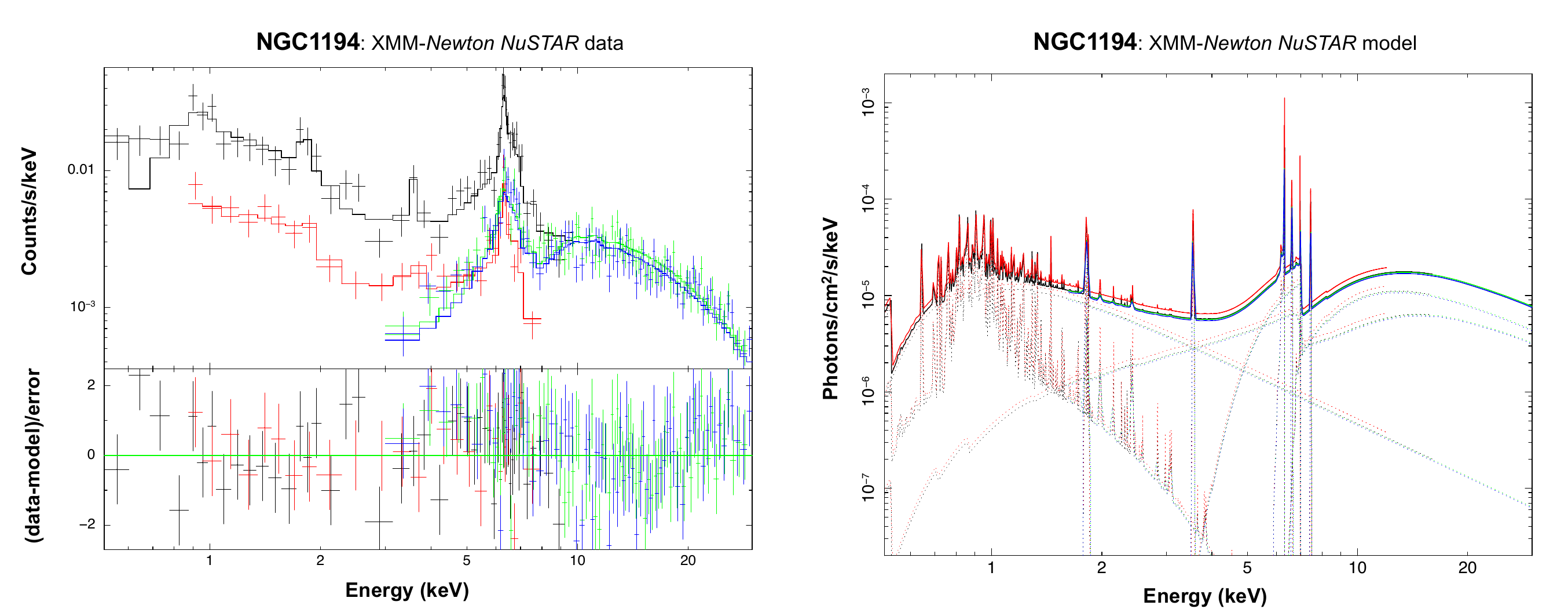}
\vglue-0.2cm
\caption[]{
{\it (Left)} 0.5--30~keV spectrum of NGC1194 (\xmm\ EPIC pn: black, EPIC MOS1$+$2: red; \nustar\ FPMA and FPMB: green and blue, respectively), fitted with the phenomenological model (top left: {\sc phabs[mekal + zgauss + zgauss + zpowerlaw + zphabs(zpowerlaw + zgauss + zgauss + zgauss + zgauss)]}) and with {\sc MYTorus}  
(bottom left: {\sc phabs(mekal + zgauss + zgauss + zgauss + zgauss + zpowerlaw + {\sc MYTorus})}), and data-model residuals in units of $\sigma$ (bottom panels of both spectra). {\it (Right)} Corresponding spectral-fitting models with the same color code. 
}
\label{fig:NGC1194_spectra}
\end{figure*}

\subsubsection{NGC1194 spectral fits}
Once verified that the \xmm\ and \nustar\ spectra provide a similar description of the NGC1194 properties and in absence of a significant flux variation in the 3--10~keV common band, being the values 
$\sim1\times10^{-12}$~\cgs\ according to both satellites, we were able to fit the full-band $\sim$0.5--30~keV spectra. 

\vglue0.2cm
\pn\textbf{Phenomenological model.} Figure~\ref{fig:NGC1194_spectra} shows the simultaneously fitted spectra (top-left panel) and the relative phenomenological transmission model (top-right panel); pn data are in black, MOS (summed) in red, \nustar\ FPMA in green and FPMB in blue. All the spectral features identified in the pn spectrum are reproduced in the MOS data, except for what may be the Ni~K$\alpha$ emission line (visible from the residuals around 7.5~keV), possibly due to the lower effective areas of the MOS detectors. 
The Galactic column density for this source is $6.0\times10^{20}$~cm$^{-2}$; also in this case an extra absorption of 
$\sim9\times10^{21}$~cm$^{-2}$ was required in the soft band. To account for the host galaxy contribution, we added a \textsc{mekal} component, with a derived temperature of kT$\sim$0.11~keV. Also included is a second power law component with spectral index fixed to $\Gamma_{1}=1.8$ to model the scattering fraction of the nuclear radiation that results equal to $\sim2\%$. In addition to the Fe~K$\alpha$ and 
Fe~K$\beta$ fluorescence lines, emission features from several other elements are evident in the pn spectrum and mostly confirmed by the summed MOS spectra (SiXIII, Ca~K$\alpha$, He-like FeXXV and Ni~K$\alpha$). 
In {\sc xspec} notation, the phenomenological model for NGC1194 is 
{\sc phabs[mekal + zgauss + zgauss + zpowerlaw + zphabs(zpowerlaw + zgauss + zgauss + zgauss + zgauss)]}.
Fluorescent lines from Ca and Si (and from other cosmically abundant elements such as C, O, Ne, Mg and S) are less observationally relevant than those from Fe and Ni due to their small fluorescence yield and because lower energy line photons have a higher probability of being absorbed before escaping the medium than higher energy photons \citep[see e.g.,][]{Matt1997}. Nevertheless, the fluorescent lines due to lighter elements can be important in ``pure-reflection" spectra (see, e.g., \citealt{Piconcelli2011}). The resulting $N_{H,2}$ is $1.0^{+0.1}_{-0.1}\times10^{24}$~cm$^{-2}$. 
The reduced $\chi^{2}$ of the fit, 281.7/214, is suggestive that further spectral complexities may be present. In fact, some positive residuals at $\sim$4-5~keV and above 15~keV are visible in Fig.~\ref{fig:NGC1194_spectra} (top-left panel, including data-model residuals, in units of $\sigma$), possibly suggesting the presence of a strong reflection component (see below for the improvement in the spectral fit using {\sc MYTorus} model). This would be in agreement with the presence of strong emission lines. 

\vglue0.2cm
\pn\textbf{MYTorus model}. Figure~\ref{fig:NGC1194_spectra} shows the {\sc MYTorus} model (bottom-right panel) fitting the full-band spectrum (bottom-left panel); all the basic spectral parameters with the relative errors are reported in Table~\ref{tab:allSources}. The resulting $N_{H,2}$ is 2.08$^{+0.21}_{-0.24}\times10^{24}$~cm$^{-2}$, hence in the Compton-thick regime. After having tested different values for the incident angle, we adopted $\theta_{obs}=65^{\circ}$, where the fit seems to converge. The absorption in the soft (Galactic+extra) has a column density of $\sim0.4\times10^{22}$~cm$^{-2}$; the thermal plasma component has a temperature $\sim$0.8~keV, and the scattered nuclear radiation results to be $\sim$1\% (consistent with the scattering reported in \citeauthor{Marchesi2018} \textcolor{blue}{2018}, hereafter M18). 
The prominent Fe~K$\alpha$ emission line is properly reproduced using the {\sc MYTorus} model. Its EW has been measured to be relatively high ($\sim$650~eV rest-frame). In addition to 
Fe~K$\alpha$ and Fe~K$\beta$, self-consistently fitted by the {\sc MYTorus} model, it was possible to describe the other emission residuals with the addition of Gaussian components, with centroid energies at $\sim1.84$~keV (Si XIII), $\sim3.60$~keV (Ca~K$\alpha$), $\sim7.54$~keV (Ni~K$\alpha$) and fixed at 6.7~keV (He-like~FeXXV). 
In {\sc xspec} notation, this model can be reported as {\sc phabs(mekal + zgauss + zgauss + zgauss + zgauss + zpowerlaw + {\sc MYTorus})}. 
The reduced $\chi^{2}$ of the fit is 257.0/219.
As suggested by the phenomenological model, the residuals at $\sim$4-5~keV and above 15~keV are well modeled by the reflection component of {\sc MYTorus} ($\Delta \chi^2\sim25$). 

NGC1194 has been subject of multiple \xray\ analyses in recent years; our results are broadly consistent with those published. In \citet[\nustar\ data]{IG2019}, \citetalias{Marchesi2018} (\xmm$+$\nustar\ data), \citet[\suzaku$+$\swift/BAT data]{Tanimoto2018} and \citetalias{Masini2016} (\nustar\ data) the source is presented as a Compton-thick AGN\footnote{The column density is below the Compton-thick threshold only in the work of \citetalias{Marchesi2018}, where NGC1194 is, however, considered as being Compton-thick within $\sim3.5\sigma$ uncertainty using the spectral curvature technique; see \citet{Koss2016} for definition and reliability of this technique}. 
More in detail, in these papers the reported column densities are  1.7$^{+0.24}_{-0.25}\times10^{24}$~cm$^{-2}$, 0.81$^{+0.09}_{-0.08}\times10^{24}$~cm$^{-2}$, 1.15$^{+0.37}_{-0.28}\times10^{24}$~cm$^{-2}$, 
and 1.4$^{+0.3}_{-0.2}\times10^{24}$~cm$^{-2}$, respectively. In all these works, the {\sc MYTorus} model has been adopted (in \citetalias{Masini2016} in its decoupled version), with the exception of \cite{Tanimoto2018}, where the N$_{\rm H}$ reported above is referred to the phenomenological, best-fitting result. Although the derived column density value from our analysis lies in the upper envelope of the published results, we note that overall the Compton-thick nature of this source is widely confirmed; the line is strong and present in all datasets, and the derived value of EW is fully consistent with literature (e.g., EW=780$^{+160}_{-140}$~eV in \citetalias{Marchesi2018}). 
For what concerns the continuum emission, both \citetalias{Marchesi2018} and \citetalias{Masini2016} found a best-fitting photon index for the primary power law of $\Gamma\sim1.5$ and $\Gamma\sim1.6$, respectively. The difference in slope between the one adopted in our paper ($\Gamma\sim1.8$) and $\Gamma\sim$1.5--1.6 can be responsible for our slightly higher value of column density given the degeneracy between the two parameters. We note, however, that once higher energy data are considered as in the work by \citet[including \swift/BAT]{Tanimoto2018}, a photon index of $\sim1.9$ is found. 

The AGN observed 2--10~keV band flux and rest-frame, absorption-corrected luminosity (see Table~\ref{tab:allSources_FLNH}) are $\sim1.0\times10^{-12}$~\cgs\ and $\sim7.0\times10^{42}$~\lum, respectively. 
While the flux is consistent with both \citetalias{Marchesi2018} and \citetalias{Masini2016} values, our \xray\ luminosity agrees with the value reported by \citetalias{Masini2016} but is a factor of $\sim$2 higher than the one in \citetalias{Marchesi2018}, most likely because of our higher correction for intrinsic absorption. 
The 10--40~keV luminosity is $\sim8.0\times10^{42}$~\lum. 


\begin{table}
	\caption[]{{\sc MYTorus} model parameters of the UGC05101, NGC1194 and NGC3079 broad-band spectral fits. We list: neutral absorption due to host galaxy in units of $10^{22}$ cm$ ^{-2} $; temperature (keV) of the thermal plasma component; neutral absorption measured in the primary power law ($ 10^{24} $ cm$ ^{-2} $); ratio of the normalization of the soft power law ($ A_{\Gamma_{2}} $) to the normalization of the primary power law ($ A_{\Gamma_{1}} $), i.e. scattering fraction; rest-frame energy (keV) and EW (eV, pn values) of the detected emission lines; reduced 
$\chi^{2}$ of the fit. Equivalent widths are computed using the phenomenological model. Errors are quoted at the 90\% confidence level for one parameter of interest.}
	\label{tab:allSources}
	\centering
	\begin{tabular}{lccc}
		\toprule \toprule
		& UGC05101 & NGC1194 & NGC3079 \\
		\midrule
		\medskip
		$N_{H,1}$ ($10^{22}$) & $ 0.82_{-0.11}^{+0.13} $ & $ 0.41_{-0.05}^{+0.07} $ & $ 0.74\pm{0.06} $ \\
		\medskip
		$kT$ & $ 0.18_{-0.03}^{+0.07}$ & $ 0.77_{-0.15}^{+0.15}$ & $ 0.19_{-0.01}^{+0.04}$ \\
		\medskip $N_{H,2}$ ($10^{24}$) & $ 1.21_{-0.17}^{+0.19}$ & $ 2.08_{-0.24}^{+0.21}$ & $ 2.35\pm{0.18}$ \\
		\medskip
		$ A_{\Gamma_{2}}/A_{\Gamma_{1}} $ & $ 0.02 $ & $ 0.01 $ & $ 0.01 $ \\
		\medskip
		$ E_{SiXIV} $ & $ 1.95_{-0.07}^{+0.07}$ & - & -\\
		\medskip
		$EW_{SiXIV}$ & $ 100_{-68}^{+61}$ & - & -\\
		\medskip
		$ E_{SiXIII} $ & - & $ 1.85_{-0.05}^{+0.10}$ & - \\
		\medskip
		$EW_{SiXIII}$ & - & $ 75_{-53}^{+64}$ & - \\
		\medskip
		$ E_{CaK_{\alpha}} $ & - & $ 3.68_{-0.12}^{+0.08}$ & -\\
		\medskip
		$EW_{CaK_{\alpha}}$ & - & $ 311_{-169}^{+147}$ & -\\
		\medskip
		$ E_{FeK_{\alpha}} $ & $ 6.58_{-0.13}^{+0.20}$ & $ 6.41_{-0.04}^{+0.02}$ & $ 6.44_{-0.04}^{+0.04}$ \\
		\medskip
		$ EW_{FeK_{\alpha}}$ & $ 153_{-153}^{+117}$ & $ 646_{-153}^{+123}$ & $ 1128_{-407}^{+642}$ \\
		\medskip
		$ E_{FeXXV} $ & - & $ 6.78_{-0.25}^{+0.56}$ & - \\
		\medskip
		$ EW_{FeXXV}$ & - & $ 66_{-59}^{+51}$ & - \\
		\medskip
		$ E_{FeK_{\beta}} $ & - & $ 7.12_{-0.10}^{+0.11}$ & - \\
		\medskip
		$ EW_{FeK_{\beta}}$ & - & $ 182_{-112}^{+196}$ & - \\
		\medskip
		$ E_{NiK_{\alpha}} $ & - & $ 7.57_{-0.11}^{+0.16}$ & - \\
		\medskip
		$ EW_{NiK_{\alpha}}$ & - & $ 258_{-200}^{+138}$ & - \\
		\medskip
		$ \chi^{2} $/dof & 92.9/89 & 257.0/219 & 306.3/246 \\
		\bottomrule \bottomrule
	\end{tabular}
\end{table}
%

\begin{figure*}
\centering
\includegraphics[width=\textwidth]{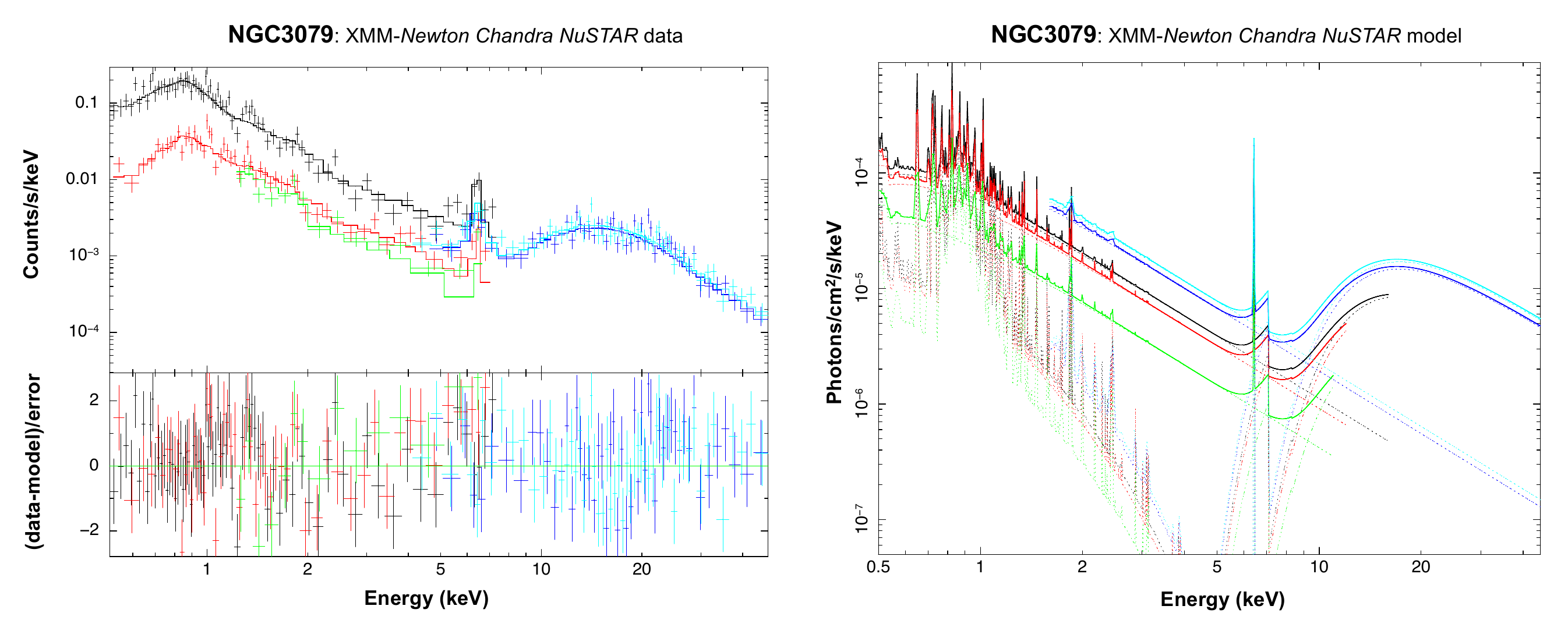}
\includegraphics[width=\textwidth]{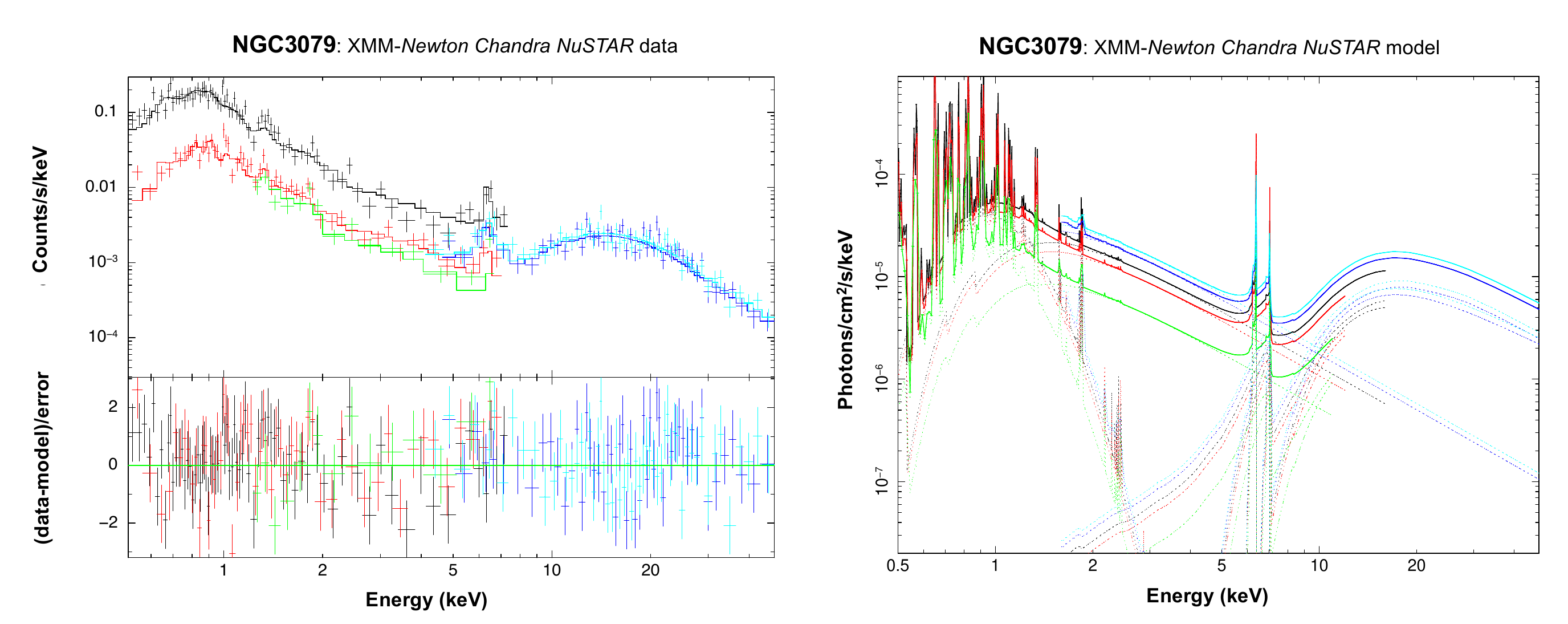}
\vglue-0.2cm
\caption[]{
{\it (Left)} 0.5--45~keV spectrum of NGC3079 (\xmm\ EPIC pn: black, EPIC MOS1$+$2: red; \chandra: green; \nustar\ FPMA and FPMB: blue and light blue, respectively), fitted with the phenomenological model (top left: {\sc phabs[mekal + zpowerlaw + zphabs(zpowerlaw + zgauss)]}) and with {\sc MYTorus} (bottom left: {\sc phabs(mekal + zpowerlaw + {\sc MYTorus})}), and data-model residuals in units of $\sigma$ (bottom panels of both spectra). {\it (Right)} Corresponding spectral-fitting models with the same color code. 
}
\label{fig:NGC3079_spectra}
\end{figure*}

\subsubsection{NGC3079 spectral fits}
Given the absence of significant flux and spectral variability among the several datasets available for NGC3079, they were all fit simultaneously in the up to $\sim$45~keV (due to the good SNR of the \nustar\ spectra) using a transmission model. 

\vglue0.2cm
\pn\textbf{Phenomenological model}. Figure~\ref{fig:NGC3079_spectra} (top-left panel) shows the fitted spectra (and data-model residuals in units of $\sigma$) and the relative model in transmission (top-right panel); pn data are shown in black, the MOS summed spectra are in red, ACIS-S data are in green, and blue and light blue show the \nustar\ FPMA and FPMB spectra, respectively. The MOS data reproduce well all the spectral features identified in the pn spectrum, although some residuals remain at the energy of the Fe~K$\alpha$ fluorescence emission line. The resulting spectral parameters derived from the \chandra\ data fitted individually have values consistent with those derived from the \xmm\ spectra, in particular the $N_{H,2}$ value measured in the primary power law. 
The Galactic absorption is $8.8\times10^{19}$~cm$^{-2}$, while the column density considered in the spectral fit and affecting the soft \xray\ emission (Galactic+extra) is $\sim8\times10^{20}$~cm$^{-2}$. A \textsc{mekal} component, with temperature kT$\sim0.64$~keV, accounts for the host galaxy emission in the soft band; the scattered fraction is $\sim1.5$\%. The neutral column density measured in the primary power law results $N_{H,2}=2.52^{+0.25}_{-0.22}\times10^{24}$~cm$^{-2}$, confirming that the source is heavily obscured. Summarizing, the best-fitting phenomenological model in {\sc xspec} notation is {\sc phabs[mekal + zpowerlaw + zphabs(zpowerlaw + zgauss)]}. 
Some residuals are still present (as suggested by the quality of the fit parameterized by $\chi^{2}$/dof=318.5/244), especially at the iron K$\alpha$ energy, possibly indicating some variations in the iron line properties between the \xmm\ vs. \nustar\ datasets. 
 
\vglue0.2cm
\pn\textbf{MYTorus model}. Figure~\ref{fig:NGC3079_spectra} (bottom-left panel) shows the full-band spectra fitted with {\sc MYTorus} model. The spectral parameters with relative errors are listed in Table~\ref{tab:allSources}. 
The absorption (Galactic+extra) in the soft band has a column density of $7.4\pm{0.6}\times10^{21}$~cm$^{-2}$; the thermal plasma component has a temperature of $\sim$0.2~keV, and the scattered nuclear radiation results to be $\sim$1\% (in agreement with \citetalias{Masini2016}). The Fe~K$\alpha$ emission line is prominent, with a corresponding (rest-frame) EW of $\sim$1.1~keV (consistent, within the errors, with the value reported by \citetalias{Marchesi2018}), indicative of heavy obscuration of the central source. Thanks to the wide band we constrained the absorption to be $N_{H,2}=2.35\pm{0.18}\times10^{24}$~cm$^{-2}$. 
This physical model, which can be reproduced within {\sc xspec} as {\sc phabs(mekal + zpowerlaw + {\sc MYTorus})}, is shown in Figure~\ref{fig:NGC3079_spectra} (bottom-right panel). 
The best value for the incident angle is close to $\theta_{obs}=90^{\circ}$, indicating an edge-on orientation of the source with respect to the line-of-sight. 
The reduced $\chi^{2}$ of the fit is 306.3/246; residuals are still present at $\sim$3--7~keV and at $\sim$20--30~keV, but likely do not affect significantly the main outcomes of our analysis. 

Also for NGC3079 we can compare the main spectral results obtained from our analysis to those reported in recent literature works. In particular, the source is reported as heavily obscured in \citet[$N_{\rm H}=3.20^{+0.54}_{-0.43}\times10^{24}$~cm$^{-2}$]{IG2019}, \citetalias{Marchesi2018} ($N_{\rm H}=2.47\pm{0.23}\times10^{24}$~cm$^{-2}$), \citetalias{Masini2016} 
($N_{\rm H}=2.5\pm{0.3}\times10^{24}$~cm$^{-2}$), and in \citet{Brightman2015}, where the {\sc TORUS} and {\sc SPHERE} models provide a column density in the range $\sim(1.5-2.7)\times10^{24}$~cm$^{-2}$, pointing towards a fully-covered picture for the obscuration in NGC3079. Furthermore, in all the cited papers the power-law photon index is $\sim1.8-1.9$, consistent with the value assumed in our spectral fitting. 

The AGN flux and absorption-corrected luminosity in the 2--10~keV band are $\sim4.3\times10^{-13}$~\cgs\ (intermediate between \citetalias{Marchesi2018} and \citetalias{Masini2016}) and $\sim9.5\times10^{41}$~\lum\ (at face value, $\sim$50\% lower than in \citetalias{Marchesi2018} and \citetalias{Masini2016}); the \xray\ luminosity in the 10--40~keV band is $\sim1.5\times10^{42}$~\lum. 

%

\begin{table*}
	\caption[]{Observed fluxes (\cgs), column densities (in units of cm$^{-2}$) and rest-frame intrinsic (i.e., absorption-corrected) luminosities (\lum). Uncertainties on fluxes and luminosities are of the order of 10--20\%, while those for column densities are reported in Table~\ref{tab:allSources} and in the main text. Both {\sc MYTorus}-derived values and those from the phenomenological model are reported for the column densities and the intrinsic luminosities.}
\label{tab:allSources_FLNH}
	\centering
	\begin{tabular}{lccccc}
		\toprule \toprule
		\multicolumn{1}{l}{SrcName} &
		\multicolumn{3}{c}{{\sc MYTorus}}  &
		\multicolumn{2}{c}{Phenomenological} \\
		\midrule
        & $F_{2-10\ keV}$ & $N_{H}$ & $L_{2-10\ keV}$ & $N_{H}$ & $L_{2-10\ keV}$ \\
		\midrule
		\medskip
		UGC05101 & $1.8\times10^{-13}$ & $1.2\times10^{24}$ & $1.2\times10^{43}$ & $1.2\times10^{24}$ & $8.6\times10^{42}$ \\
		\medskip
		NGC1194  & $1.0\times10^{-12}$ & $2.1\times10^{24}$ & $7.0\times10^{42}$ & $1.0\times10^{24}$ & $4.5\times10^{42}$ \\
		\medskip 
        NGC3079  & $4.3\times10^{-13}$ & $2.4\times10^{24}$ & $9.5\times10^{41}$ & $2.5\times10^{24}$ & $3.9\times10^{41}$ \\
		\bottomrule \bottomrule
	\end{tabular}
\end{table*}
\section{Main X-ray spectral fitting results}
\label{sec:Main broad-band analysis results}
The wide bandpass, obtained fitting simultaneously \nustar\ spectra along with either \xmm\ or \chandra\ (or both) spectra, allowed us to gain new and more solid insights into the \xray\ spectral properties of the 12MGS sources presented in this paper. We focused on an accurate determination of the column density parameter taking advantage of the availability of broad-band data. The resulting measured absorption column densities, in units of $10^{24}$~cm$^{-2}$, are $1.21^{+0.19}_{-0.17}$, 
$2.08^{+0.21}_{-0.24}$ and $2.35\pm{0.18}$ for UGC05101, NGC1194 and NGC3079, respectively, therefore in the Compton-thick regime. This confirms that the original selection of these sources (see Section~\ref{sec:selection_oursubsample}) was correct. The consistency of our results will be discussed using the \cite{Gandhi2009} relation in $\S$\ref{discussion}. The rest-frame EWs of the K$\alpha$ iron fluorescent emission line are about 150~eV, 650~eV and 1.1~keV, respectively. These values are highly suggestive of heavy obscuration (at least for NGC1194 and NGC3079). For UGC05101, even if all spectral models suggest a Compton-thick nature of the source, the iron line results to be weak, consistent with the results obtained by \cite{Oda2017} and \cite{Iwasawa2011}. A possible explanation for the low EW of the line could be either a low metallicity or a low covering factor for the torus. To check whether the latter hypothesis is viable, we fitted the data using the {\sc BORUS02} model \citep{Balokovic2018}, where the covering factor of the obscuring medium is a free parameter of the fit. The fit is worse ($\chi^{2}$/dof=119/87) than the one obtained with {\sc MYTorus}, but overall the main components described above and the derived spectral parameters (including the column density of the obscuring medium) are confirmed. The resulting covering factor of the torus is constrained to be $<0.28$ (where unity means a fully-covered source), and the torus appears to be observed at an inclination angle close to 90\arcdeg\ (i.e., edge-on). Hence, a low covering factor for the torus can provide a reasonable explanation for the low EW of the iron emission line reported above. 

As exercise meant to determine how much one can be wrong in using a phenomenological model (a transmission model in this case) and as {\it a posteriori} verification, we compared the column densities derived using the two tested models, i.e. {\sc MYTorus} and the phenomenological model. Table~\ref{tab:allSources_FLNH} lists the values. They are broadly consistent and, overall, all indicate that the sources suffer from heavy obscuration.  
The intrinsic (i.e. corrected for the absorption) AGN rest-frame luminosity in the 2--10~keV band is $1.2\times10^{43}$, 
$7.0\times10^{42}$ and $9.5\times10^{41}$~\lum\ in case of UGC05101, NGC1194 and NGC3079, respectively, all obtained using the {\sc MYTorus} model. We investigated the 2--10~keV luminosity of the AGN using also the phenomenological transmission model to evaluate how much model-dependent the results are. Table~\ref{tab:allSources_FLNH} lists the values derived from the different spectral models. It results that {\sc MYTorus} systematically provides higher luminosity values with respect to the phenomenological model, without necessarily finding higher $N_{H}$ values. In the most extreme case (NGC3079), the value inferred from the phenomenological model is $\sim$40\% of the {\sc MYTorus} one.

\section{Discussion and conclusions}
\label{discussion}
In this paper we have focused on the search and detailed \xray\ spectral analysis of heavily obscured AGN candidates. The sources have been selected from the subsample of the local 12MGS considered by \citetalias{Gruppioni2016} to trace black hole accretion by means of SED decomposition. 
\citetalias{Gruppioni2016} compared the AGN bolometric luminosity ($L_{bol}^{AGN}$) derived from the SED decomposition to the same quantity obtained by means of different methods, among which the 2--10~keV luminosity converted into a bolometric luminosity using a bolometric correction. From the $L_{bol}^{AGN}(IR)/L_{bol}^{AGN}(X)$ ratios, a difference up to three orders of magnitude resulted for some sources, suggesting that the original intrinsic \xray\ luminosity, although formally corrected for the obscuration, could be underestimated. In particular, in the present work we have focused on three of the most extreme of such sources, showing a significant difference with respect to the 1:1 relation and for which data from at least two satellites were available, one of these being \nustar, complementary to \xmm\ and/or \chandra. 
\nustar\ extends the spectral band up to high energies ($\sim$30--45~keV), whereas \xmm\ and \chandra\ provide spectral coverage down to 0.5~keV and good spectral resolution ($\sim$150~eV at energy 6.4~keV) and photon statistics below 10~keV. This approach is necessary to constrain all the spectral components in an appropriate way. The resulting selected sources are UGC05101, NGC1194 and NGC3079; for these sources we provided an accurate analysis of the \xray\ spectra. 
We aimed at determining their obscuration and, hence, their true intrinsic power, 
by applying the physically motivated {\sc MYTorus} model to \xray\ data. 

The \xray\ spectral properties inferred from our broad-band analysis are highly suggestive of heavy obscuration. In particular, the measured absorption column densities, in units of $10^{24}$~cm$^{-2}$, are $\sim1.2$, $\sim2.1$ and $\sim2.4$ for UGC05101, NGC1194 and NGC3079, respectively, with typical errors of the order of 12\% (at the 90\% confidence level), therefore in the Compton-thick regime. The large measured $N_{H}$ values are consistent with the prominent dust absorption features attributed to silicates present in the \spitzer-IRS spectra of these sources (Fig.~\ref{fig:IRSspectra}), and with previous analyses (e.g., \citealt{Oda2017}; \citetalias{Masini2016, Marchesi2018}).


Furthermore, compared to other literature works dealing with a broad-band analysis as the one presented in this paper, we have strictly linked the \xray\ properties of the sources with those derived in the mid-IR via SED-fitting decomposition (\citetalias{Gruppioni2016}). 
Of the sources listed in Tab.~1, all the three analyzed here ($\sim8$\%) are recognized as Compton-thick once the \nustar\ data are taken into account. Interestingly, \citetalias{Marchesi2018} found a fraction of $\sim40\%$ of hard \xray\ selected sources (from the Swift-Burst Alert Telescope (BAT) 100 month catalog) for which the addition of \nustar\ data has the opposite effect, turning their classification from Compton-thick to Compton-thin, thus underlying again the importance of a broad-band approach for local obscured AGN.

To finally verify the intrinsic \xray\ power derived from the spectral analysis, we compared the absorption-corrected 2--10~keV luminosities (adopting the {\sc MYTorus} modeling) with those derived from the mid-IR band using the empirical $L_{MIR}-L_{X}$ relation by \cite{Gandhi2009}; see also \cite{Asmus2015}. 
The mid-IR (12.3\micron) luminosity for the AGN component alone is derived directly from the SED fitting carried out by \citetalias{Gruppioni2016} and can be considered an accurate proxy of the intrinsic AGN power, since any primary continuum (i.e., disc emission) that is absorbed must ultimately come out at these wavelengths after being thermally reprocessed by the torus \citep[e.g.,][]{Lutz2004, Maiolino2007}. 
We therefore expect that the \xray\ luminosities obtained from \xray\ fitting and those derived from the mid-IR AGN emission are in good agreement. 
Table~\ref{tab:Gandhi+09} lists the $L_{MIR}$ values along with the comparison between $L_{[2-10\ keV]}$ derived from the \cite{Gandhi2009} relation and those obtained from the \xray\ spectral analysis. Overall, the values are in good agreement, further supporting that the estimate of the column densities from the \xray\ spectral analysis carried out in this work is likely correct.
\begin{table}
	\centering
	\caption[]{Mid-IR (12.3\micron) luminosities obtained from the SED-fitting decomposition (\citetalias{Gruppioni2016}) and 2--10~keV luminosities derived from the mid-IR-\xray\ relation of \cite{Gandhi2009}. The latter values are compared to those obtained using the {\sc MYTorus} model. All luminosities are in units of \lum; errors are of the order of 10--20\%.}
	\label{tab:Gandhi+09}
	\begin{tabular}{lccc}
		\hline \hline
		\multirow{3}{*}{Name} &
		\multirow{3}{*}{$L_{MIR}$} &
		\multicolumn{2}{c}{$L_{[2-10~keV]}$} \\
		\cmidrule(lr){3-4}
		& & derived & measured \\
		\hline
		UGC05101 & $ 3.8\times10^{43} $ & $ 2.3\times10^{43} $ & $ 1.2\times10^{43} $ \\
		NGC1194 & $ 1.3\times10^{43} $ & $ 8.4\times10^{42} $ & $ 7.0\times10^{42} $ \\
		NGC3079 & $ 1.1\times10^{42} $ & $ 9.5\times10^{41} $ & $ 9.5\times10^{41} $ \\
		\hline \hline
	\end{tabular}
\end{table}
%
\begin{figure}
	\centering
    \includegraphics[clip=true,width=\columnwidth]{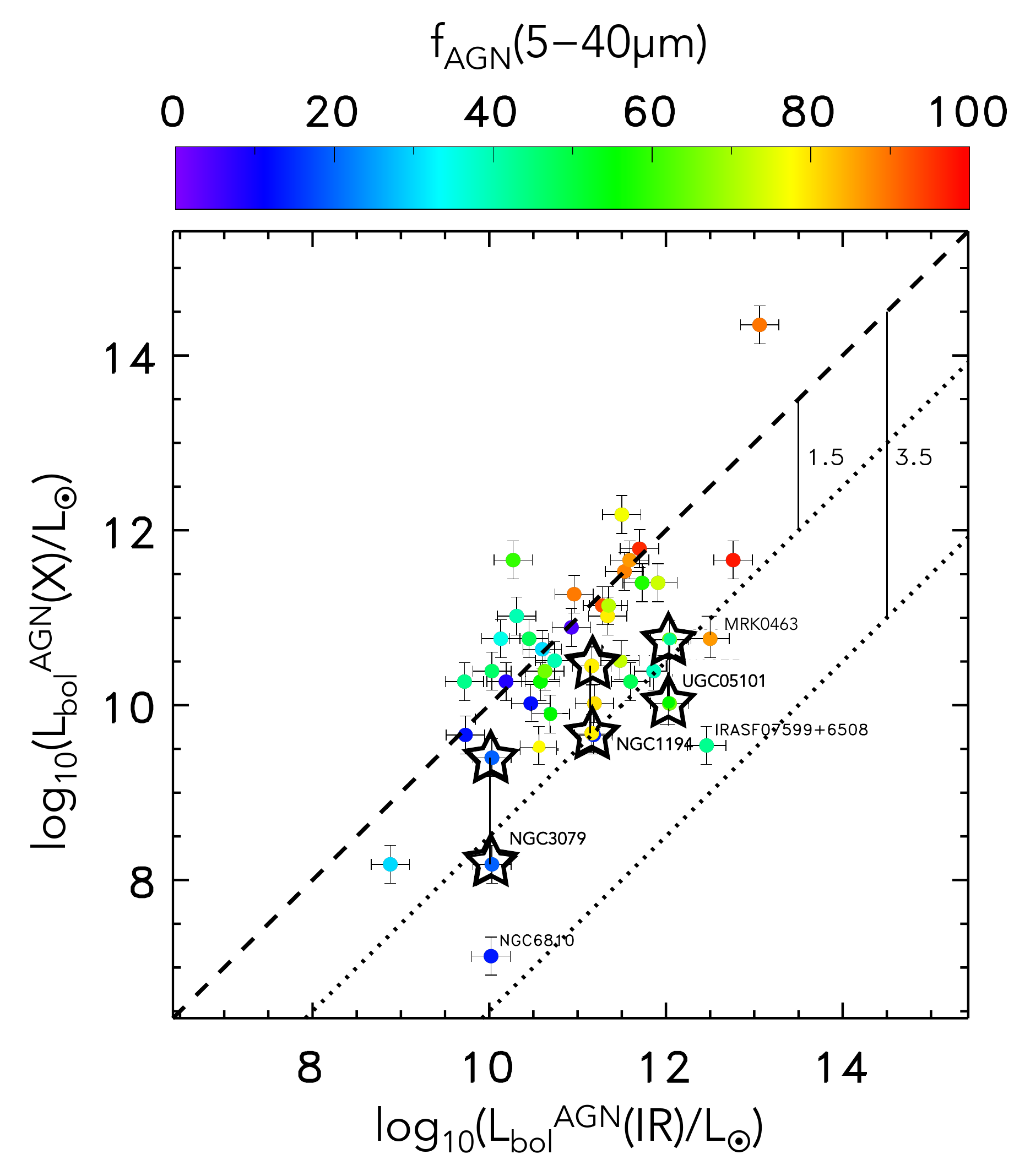}
    \vglue-0.15cm
	\caption[]{AGN bolometric luminosity obtained from the best-fitting torus model \citepalias{Gruppioni2016} versus AGN bolometric luminosity that has been computed from the 2--10~keV luminosity derived from our spectral analysis applying uniformly the \citetalias{Marconi2004} bolometric correction. 
	The current and the previous positions of the three sources selected for our analysis in the starting plot (Fig.~\ref{fig:Fig6Gruppioni}) are indicated by star symbols, connected by a vertical straight line.}
	\label{fig:fig6Marconi_ourValues}
\end{figure}

%
To conclude our work, we replotted the original relation (Fig.~\ref{fig:Fig6Gruppioni}) using the AGN bolometric luminosity $L_{bol}^{AGN}(X)$ that we have computed from the 2--10~keV luminosity derived from our spectral analysis applying the \citetalias{Marconi2004} bolometric correction. The new source distribution is shown in Figure~\ref{fig:fig6Marconi_ourValues}. The examined sources that showed the most striking separation from the linear relation (e.g., $1.5-3.0$ times lower in log) now lie much closer to the 1:1 relation, within the range of 1.5 (see the star symbols connected by a straight vertical line in Fig.~\ref{fig:fig6Marconi_ourValues}). 

The selection of AGN with very low observed \xray\ luminosity to 
$L_{IR}$ ratios \citep[e.g.,][Dalla Mura et al. in preparation]{Lanzuisi2015a} has been extensively used recently to identify Compton-thick sources, even at high redshift, although for a more complete source characterization (e.g., by using the most up-to-date torus models) good-quality \xray\ data are needed. This work can pave the way for future-mission studies as an obscured AGN selection method like the one adopted in this paper (not considering the complication of recomputing the $k_{bol}$) could be effective in perspective of the \xray\ satellite \erosita\ \citep{Merloni2012} in combination with either WISE or the infrared facility \spica\ (e.g., \citealt{Gruppioni2017, Roelfsema2018}) selected as a new-mission concept study by ESA for M5.

\section*{Acknowledgements}
Tha authors thank the referee for her/his useful comments and suggestions, which helped improving the quality of the paper. 
The authors acknowledge financial support from the Italian Space Agency (ASI) under the contracts ASI-INAF I/037/12/0 and ASI-INAF n.2017-14-H.0. 





\begin{thebibliography}{99}

\bibitem[\protect\citeauthoryear{Ajello et al.}{2012}]{Ajello2012}
Ajello M., Alexander D.~M., Greiner J. et al., 2012, ApJ, 749, 21

\bibitem[\protect\citeauthoryear{Akylas et al.}{2012}]{Akylas2012} 
Akylas A., Georgakakis A., Georgantopoulos I., Brightman M., Nandra K., 2012, A\&A, 546, A98 

\bibitem[\protect\citeauthoryear{Arnaud}{1996}]{Arnaud1996} 
Arnaud K.~A., 1996, ASPC, 101, 17 

\bibitem[\protect\citeauthoryear{Asmus et al.}{2015}]{Asmus2015}
Asmus D., Gandhi P., H{\"o}nig, S.~F. et al., 2015, MNRAS, 454, 766

\bibitem[\protect\citeauthoryear{Avni}{1976}]{Avni1976}
Avni Y., 1976, ApJ, 210, 642

\bibitem[\protect\citeauthoryear{Balokovi{\'c} et al.}{2018}]{Balokovic2018} 
Balokovi{\'c} M., et al., 2018, ApJ, 854, 42 


\bibitem[\protect\citeauthoryear{Beckmann et al.}{2009}]{Beckmann2009}
Beckmann V., Soldi S., Ricci C. et al., 2009, A\&A, 505, 417

\bibitem[\protect\citeauthoryear{Bianchi et al.}{2008}]{Bianchi2008} 
Bianchi S., Chiaberge M., Piconcelli E., Guainazzi M., Matt G., 2008, MNRAS, 386, 105 

\bibitem[\protect\citeauthoryear{Brandt \& Alexander}{2015}]{Brandt2015}
Brandt W.~N., Alexander D.~M., 2015, A\&AR, 23, 1

\bibitem[\protect\citeauthoryear{Brightman \& Nandra}{2011a}]{BrightmanNandra2011a}
Brightman M., Nandra K., 2011a, MNRAS, 413, 1206 (BN11a)

\bibitem[\protect\citeauthoryear{Brightman \& Nandra}{2011b}]{BrightmanNandra2011b}
Brightman M., Nandra K., 2011b, MNRAS, 414, 3084

\bibitem[\protect\citeauthoryear{Brightman et al.}{2015}]{Brightman2015} 
Brightman M., et al., 2015, ApJ, 805, 41 


\bibitem[\protect\citeauthoryear{Cappi et al.}{2006}]{Cappi2006}
Cappi M., Panessa F., Bassani L. et al., 2006, A\&A, 446, 459

\bibitem[\protect\citeauthoryear{Castangia et al.}{2013}]{Castangia2013} Castangia P., Panessa F., Henkel C., Kadler M., Tarchi A., 2013, MNRAS, 436, 3388 


\bibitem[\protect\citeauthoryear{Del Moro et al.}{2017}]{DelMoro2017} 
Del Moro A., et al., 2017, ApJ, 849, 57 

\bibitem[\protect\citeauthoryear{Feltre et al.}{2012}]{Feltre2012}
Feltre A., Hatziminaoglou E., Fritz J. et al., 2012, MNRAS, 426, 120

\bibitem[\protect\citeauthoryear{Fritz et al.}{2006}]{Fritz2006}
Fritz J., Franceschini A., Hatziminaoglou E., 2006, MNRAS, 366, 767

\bibitem[\protect\citeauthoryear{Gandhi et al.}{2009}]{Gandhi2009}
Gandhi P., Horst H., Smette A. et al., 2009, A\&A, 502, 457

\bibitem[\protect\citeauthoryear{Georgantopoulos \& Akylas}{2019}]{IG2019} 
Georgantopoulos I., Akylas A., 2019, A\&A, 621, A28 

\bibitem[\protect\citeauthoryear{Gilli et al.}{2007}]{Gilli2007}
Gilli R., Comastri A., Hasinger G., 2007, A\&A, 463, 79

\bibitem[\protect\citeauthoryear{Gilli}{2013}]{Gilli2013} 
Gilli R., 2013, MmSAI, 84, 647 

\bibitem[\protect\citeauthoryear{Gilli et al.}{2014}]{Gilli2014} 
Gilli R., et al., 2014, A\&A, 562, A67 

\bibitem[\protect\citeauthoryear{Greenhill, Tilak, \& Madejski}{2008}]{Greenhill2008} 
Greenhill L.~J., Tilak A., Madejski G., 2008, ApJ, 686, L13 

\bibitem[\protect\citeauthoryear{Gruppioni et al.}{2016}]{Gruppioni2016}
Gruppioni C., Berta S., Spinoglio L. et al., 2016, MNRAS, 458, 4297 (G16)

\bibitem[\protect\citeauthoryear{Gruppioni et al.}{2017}]{Gruppioni2017} 
Gruppioni C., et al., 2017, PASA, 34, e055 

\bibitem[\protect\citeauthoryear{Harrison et al.}{2013}]{Harrison2013}
Harrison F.~A., Craig W.~W., Christensen F.~E. et al., 2013, ApJ, 770, 103

\bibitem[\protect\citeauthoryear{Harrison et al.}{2016}]{Harrison2016}
Harrison F.~A., Aird J., Civano F. et al., \ 2016, \apj, 831, 185  

\bibitem[\protect\citeauthoryear{Hickox \& Markevitch}{2006}]{Hickox2006}
Hickox R.~C., Markevitch M., 2006, ApJ, 645, 95


\bibitem[\protect\citeauthoryear{Ikeda, Awaki, \& Terashima}{2009}]{Ikeda2009} 
Ikeda S., Awaki H., Terashima Y., 2009, ApJ, 692, 608 

\bibitem[\protect\citeauthoryear{Imanishi et al.}{2003}]{Imanishi2003} 
Imanishi M., Terashima Y., Anabuki N., Nakagawa T., 2003, ApJ, 596, L167 

\bibitem[\protect\citeauthoryear{Iwasawa et al.}{2011}]{Iwasawa2011} 
Iwasawa K., et al., 2011, A\&A, 529, A106 

\bibitem[\protect\citeauthoryear{Kalberla et al.}{2005}]{Kalberla2005}
Kalberla P.~M.~W., Burton W.~B., Hartmann D. et al., 2005, A\&A, 440, 775

\bibitem[\protect\citeauthoryear{Kondratko, Greenhill, \& Moran}{2005}]{Kondratko2005} 
Kondratko P.~T., Greenhill L.~J., Moran J.~M., 2005, ApJ, 618, 618 

\bibitem[\protect\citeauthoryear{Koss et al.}{2016}]{Koss2016} 
Koss M.~J., et al., 2016, ApJ, 825, 85 

\bibitem[\protect\citeauthoryear{Krivonos et al.}{2007}]{Krivonos2007}
Krivonos R., Revnivtsev M., Lutovinov A. et al., 2007, A\&A, 475, 775

\bibitem[\protect\citeauthoryear{Lansbury et al.}{2017a}]{Lansbury2017a} 
Lansbury G.~B., et al., 2017a, ApJ, 836, 99 

\bibitem[\protect\citeauthoryear{Lansbury et al.}{2017b}]{Lansbury2017b} 
Lansbury G.~B., et al., 2017b, ApJ, 846, 20 

\bibitem[\protect\citeauthoryear{Lanzuisi et al.}{2015a}]{Lanzuisi2015a}
Lanzuisi G., Perna M., Delvecchio I. et al., 2015a, A\&A, 578, A120

\bibitem[\protect\citeauthoryear{Lanzuisi, et al.}{2015b}]{Lanzuisi2015b}
Lanzuisi G., et al., 2015b, A{\&}A, 573, A137

\bibitem[\protect\citeauthoryear{Leiter et al.}{2018}]{Leiter2018} 
Leiter K., et al., 2018, IAUS, 336, 141 

\bibitem[\protect\citeauthoryear{Luo et al.}{2014}]{Luo2014}
Luo B., Brandt W.~N.. Alexander D.~M. et al., 2014, \apj, 794, 70

\bibitem[\protect\citeauthoryear{Lusso et al.}{2012}]{Lusso2012}
Lusso E., Comastri A., Simmons B.~D. et al., 2012, MNRAS, 425, 623

\bibitem[\protect\citeauthoryear{Lutz et al.}{2004}]{Lutz2004} 
Lutz D., Maiolino R., Spoon H.~W.~W., Moorwood A.~F.~M., 2004, A\&A, 418, 465 



\bibitem[\protect\citeauthoryear{Maiolino et al.}{2007}]{Maiolino2007}
Maiolino R., Shemmer O., Imanishi M. et al., 2007, A\&A, 468, 979

\bibitem[\protect\citeauthoryear{Marchesi et al.}{2018}]{Marchesi2018}
Marchesi S., Ajello M., Marcotulli L., Comastri A., Lanzuisi G., Vignali C., 2018, ApJ, 854, 49 (M18)

\bibitem[\protect\citeauthoryear{Marchesi et al.}{2019}]{Marchesi2019} 
Marchesi S., et al., 2019, ApJ, 872, 8 

\bibitem[\protect\citeauthoryear{Marconi et al.}{2004}]{Marconi2004}
Marconi A., Risaliti G., Gilli R. et al., 2004, MNRAS, 351, 169 (M04)

\bibitem[\protect\citeauthoryear{Martocchia et al.}{2017}]{Martocchia2017} 
Martocchia S., et al., 2017, A\&A, 608, A51 

\bibitem[\protect\citeauthoryear{Masini et al.}{2016}]{Masini2016}
Masini A., Comastri A., Balokovi{\'c} M. et al., 2016, \aap, 589, A59 (M16)

\bibitem[\protect\citeauthoryear{Matt et al.}{1997}]{Matt1997}
Matt G., Fabian A.~C., Reynolds C.~S., 1997, MNRAS, 289, 175

\bibitem[\protect\citeauthoryear{Merloni et al.}{2012}]{Merloni2012} 
Merloni A., et al., 2012, MPE document. Edited by S. Allen. G. Hasinger and K. Nandra (arXiv:1209.3114)

\bibitem[\protect\citeauthoryear{Murphy \& Yaqoob}{2009}]{Murphy2009}
Murphy K.~D., Yaqoob T., 2009, MNRAS, 397, 1549

\bibitem[\protect\citeauthoryear{Oda et al.}{2017}]{Oda2017}
Oda S., Tanimoto A., Ueda Y. et al., 2017, ApJ, 835, 179

\bibitem[\protect\citeauthoryear{Pesce et al.}{2015}]{Pesce2015} 
Pesce D.~W., Braatz J.~A., Condon J.~J., Gao F., Henkel C., Litzinger E., Lo K.~Y., Reid M.~J., 2015, ApJ, 810, 65 

\bibitem[\protect\citeauthoryear{Piconcelli et al.}{2005}]{Piconcelli2005}
Piconcelli E., Jimenez-Bail{\'o}n E., Guainazzi M. et al., 2005, A\&A, 432, 15

\bibitem[\protect\citeauthoryear{Piconcelli et al.}{2011}]{Piconcelli2011}
Piconcelli E., Bianchi S., Vignali C., Jim{\'e}nez-Bail{\'o}n E., Fiore, F.,  2011, \aap, 534, A126 

\bibitem[\protect\citeauthoryear{Pozzi et al.}{2012}]{Pozzi2012} 
Pozzi F., et al., 2012, MNRAS, 423, 1909 


\bibitem[\protect\citeauthoryear{Ricci et al.}{2015}]{Ricci2015} 
Ricci C., Ueda Y., Koss M.~J., Trakhtenbrot B., Bauer F.~E., Gandhi P., 2015, ApJ, 815, L13 

\bibitem[\protect\citeauthoryear{Roelfsema et al.}{2018}]{Roelfsema2018} 
Roelfsema P.~R., et al., 2018, PASA, 35, e030 

\bibitem[\protect\citeauthoryear{Rush et al.}{1993}]{Rush1993}
Rush B., Malkan M.~A., Spinoglio L., 1993, ApJS, 89, 1


\bibitem[\protect\citeauthoryear{Severgnini et al.}{2015}]{Severgnini2015}
Severgnini P., Ballo L., Braito V. et al., 2015, MNRAS, 453, 3611

\bibitem[\protect\citeauthoryear{Shi et al.}{2013}]{Shi2013}
Shi Y., Helou G., Armus L., 2013, ApJ, 777, 6

\bibitem[\protect\citeauthoryear{Strickland}{2007}]{Strickland2007} 
Strickland D.~K., 2007, MNRAS, 376, 523 

\bibitem[\protect\citeauthoryear{Tanimoto et al.}{2018}]{Tanimoto2018} 
Tanimoto A., Ueda Y., Kawamuro T., Ricci C., Awaki H., Terashima Y., 2018, ApJ, 853, 146 

\bibitem[\protect\citeauthoryear{Tommasin et al.}{2008}]{Tommasin2008}
Tommasin S., Spinoglio L., Malkan M.~A. et al., 2008, ApJ, 676, 836

\bibitem[\protect\citeauthoryear{Tommasin et al.}{2010}]{Tommasin2010}
Tommasin S., Spinoglio L., Malkan M.~A. et al., 2010, ApJ, 709, 1257

\bibitem[\protect\citeauthoryear{Tueller et al.}{2008}]{Tueller2008}
Tueller J., Mushotzky R.~F., Barthelmy S. et al., 2008, ApJ, 681, 113

\bibitem[\protect\citeauthoryear{Turner et al.}{1997}]{Turner1997}
Turner T.~J., George I.~M., Nandra K., Mushotzky R.~F., 1997, ApJS, 113, 23

\bibitem[\protect\citeauthoryear{Vasudevan et al.}{2013}]{Vasudevan2013}
Vasudevan R.~V., Mushotzky R.~F., Gandhi P., 2013, ApJL, 770, L37


\bibitem[\protect\citeauthoryear{Vignali et al.}{2009}]{Vignali2009} 
Vignali C., et al., 2009, MNRAS, 395, 2189 

\bibitem[\protect\citeauthoryear{Vignali et al.}{2010}]{Vignali2010}
Vignali C., Alexander D.~M., Gilli R., Pozzi F., 2010, MNRAS, 404, 48

\bibitem[\protect\citeauthoryear{Vignali et al.}{2014}]{Vignali2014}
Vignali C., Mignoli M., Gilli R. et al., 2014, A\&A, 571, A34

\bibitem[\protect\citeauthoryear{Yamada et al.}{2018}]{Yamada2018} 
Yamada S., Ueda Y., Oda S., Tanimoto A., Imanishi M., Terashima Y., Ricci C., 2018, ApJ, 858, 106 

\bibitem[\protect\citeauthoryear{Zappacosta et al.}{2018}]{Zappacosta2018} 
Zappacosta L., et al., 2018, ApJ, 854, 33 

\end{thebibliography}





%
%


\bsp	
\label{lastpage}
\end{document}